\documentclass[fleqn,10pt]{wlscirep}
\usepackage[utf8]{inputenc}
\usepackage[T1]{fontenc}
\usepackage{lineno}
\usepackage{xcolor}
\usepackage{amsmath}
\usepackage{float}
\title{Waveguide Holography: Towards True 3D Holographic Glasses}

\author[1,+,*]{Changwon Jang}
\author[1,+]{Kiseung Bang}
\author[2]{Minseok Chae}
\author[2]{Byoungho Lee}
\author[1]{Douglas Lanman}
\affil[1]{Reality Labs Research, Meta, Redmond, 98052, WA, United States}
\affil[2]{Seoul National University, Seoul, 08826, Republic of Korea}
\affil[*]{changwon.jang@fb.com}
\affil[+]{these authors contributed equally to this work}

\begin{abstract}
We present a novel near-eye display concept which consists of a waveguide combiner, a spatial light modulator, and a laser light source. The proposed system can display true 3D holographic images through see-through pupil-replicating waveguide combiner as well as providing a large eye-box. By modeling the coherent light interaction inside of the waveguide combiner, we demonstrate that the output wavefront from the waveguide can be controlled by modulating the wavefront of input light using a spatial light modulator. This new possibility allows combining a holographic display, which is considered as the ultimate 3D display technology, with the state-of-the-art pupil replicating waveguides, enabling the path towards true 3D holographic augmented reality glasses.

\end{abstract}
\begin{document}

\flushbottom
\maketitle
%
%
\thispagestyle{empty}

\section*{Introduction}


    
    
    

Near-eye display technology is evolving rapidly along with the pursuit of metaverse. Especially for augmented reality (AR)~\cite{Azuma1997}, various see-through near-eye display architectures have been invented and explored in the recent decades such as birdbath type displays, curved mirror type displays, retinal projection displays, and pin mirror displays~\cite{Kress2013,bleereview}. Among the plethora architectures, pupil-replicating waveguide image combiner remains the strongest candidate for augmented reality glasses in the industry because of its compact form factor. On the other hands, there have been a lot of efforts to realize 3D holographic displays that provide realistic visual experience. In this work, we propose a novel display concept that combines the advantages of both pupil replicating waveguides and holographic displays, enabling the path towards true 3D holographic AR glasses.

\subsection*{Waveguide image combiner}
As a near-eye display application, waveguide image combiner refers to a thin transparent slab that guides the light as a total internal reflection (TIR) mode, to deliver to the user's eye. Waveguides can be designed with different types of light coupling method and materials. Geometric waveguides use partial reflective surfaces inside the slab to re-direct and extract the light from the waveguide\cite{levola20067, amitai2010substrate, cheng2014design, xu2019methods}. Diffractive waveguides may utilize surface relief grating, volume Bragg grating, polarization grating, and meta-surface or geometric phase element as in/out-couplers.\cite{kress201711,Gu_pvhwaveguide}
The optical path can be secured in the waveguide without being obstructed, while no bulky projector or imaging optics are needed to be placed in front of user’s eye. The image projector of waveguide display is typically located at the temple side with infinity corrected lens, providing high resolution images. The most unique advantage of waveguide is its étendue expansion ability via pupil replication process\cite{ayras2009exit}. Pupil-replication provides sufficient eyebox and large field of view while many other architectures suffer from their trade-off relation imposed by limited étendue. The compact form factor and pupil replication ability are the main advantages that make waveguide image combiner as an industry norm these days\cite{kress2020optical}.

Despite the unique advantages of waveguide image combiners, there are some limitations. First, it is challenging to achieve enough brightness using conventional light sources such as micro light-emitting diode (LED)\cite{erickson2020exploring}. Typically, overall throughput efficiency from light source to retina is extremely low since waveguide image combiner is intrinsically a leaky system. Laser light source can be much more power efficient than micro LED, however, the coherent light interaction causes several artifacts and degrades the image quality. Second, waveguide display can only display a single depth, normally infinity conjugate image. If finite-conjugate image is projected to the waveguide, the pupil replication process produces copies of different optical path and aberration that create severe ghost noise. However, it has long been desired that AR display to provide accommodation cue for realistic and comfort visual experience\cite{chang2020toward}. Dual/multi-imaging planes waveguide architecture have been studied\cite{shi2022design,Yoo:19}, but they come with the cost of degraded performance, bulkier form factor and additional hardware constraints.

\begin{figure}[H]
\centering
\includegraphics[width=\linewidth]{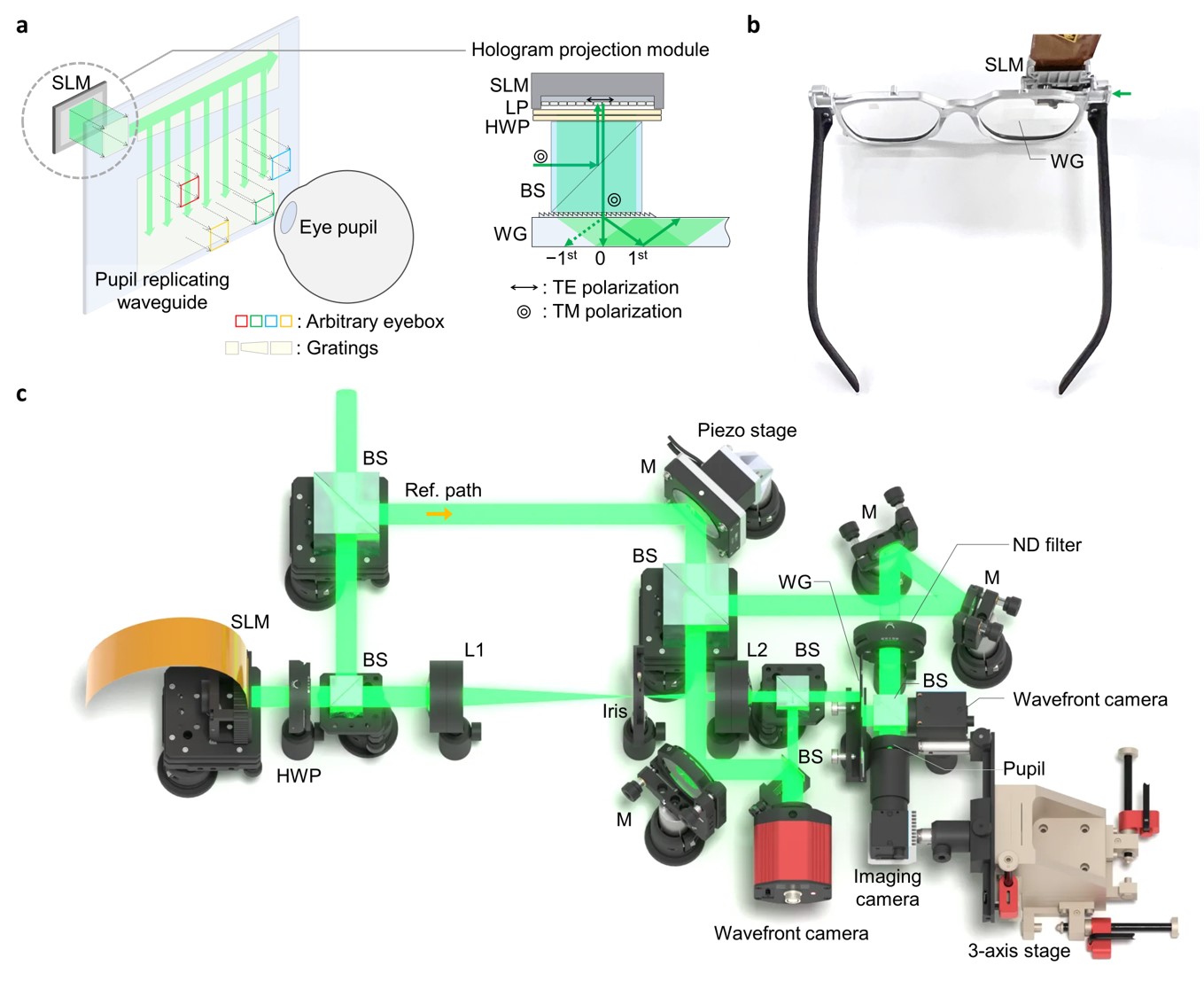}
\caption{\textbf{a} Conceptual illustration of waveguide holography architecture using the pupil replicating waveguide (WG). The hologram projection module consists of SLM, a linear polarizer (LP), a half wave plate (HWP), and beam splitter (BS) for illumination path. Apertures with different colors indicates software-steered eyebox that can be formed in the arbitrary 3D space at the out-coupler side, fully exploiting the étendue expansion capability of the waveguide. \textbf{b} A photograph of the compact prototype of waveguide holography architecture for the proof-of-concept. The SLM size can be further reduced since the active area is only as large as the input coupler size, which is 20\% of the total SLM area. \textbf{c} A benchtop prototype built for design iteration and benchmarking the performance. L1 and L2: lenses for 4-$f$ imaging system, M: mirror. See Method for the details of the system implementation.}
\label{fig:architecture}
\end{figure}

\subsection*{Holographic displays}
Meanwhile, holographic display is believed to be the ultimate 3D display technology which can provide accommodations by modulating the wavefront of coherent light using spatial light modulators (SLMs)\cite{chang2020toward}. It also offers unique benefits such as aberration-free high resolution images, per-pixel depth control, vision correction functionality\cite{nam2020aberration, kim2021vision, Kavakli:21}, as well as large color gamut. Recently, a lot of progress has been made in the field of computer-generated hologram (CGH) rendering, getting more attentions from the industry \cite{Maimone:2017, Shi:2021, Choi:2021,Peng:2020:NeuralHolography,aksit2022perceptually,gracekuo,chakravarthula2019wirtinger,jpegpleno:21,yang2022perceptually,park2019ultrathin,samsung20,peng2021partiallycoherent,kim2022holographicglasses,Matsushima:05,Kaczorowski:16}. Many conventional issues of holographic displays such as speckle, image quality issue, and heavy computation load are demonstrated to be solved by aid of better CGH rendering model and improved computation power of recent graphics processing units (GPU)\cite{chen2015improved, shimobaba2015review, lee2020deep,mengu2016non }.  
However, the architecture of near-eye holographic display for augmented reality remains as unsolved problem because of the étendue limitation\cite{Kozacki:12, gracekuo}, and it is even more difficult to achieve a glasses form factor. The retinal projection type designs have been explored with holographic projector at the temple side that projects the hologram via oblique free-space projection to the eyepiece combiner\cite{Maimone:2017,Jang:2018,Park:18}. However, such configuration has limited space and angular bandwidth to transmit enough étendue to the eyepiece, even with mechanical pupil steering\cite{Jang:2018,Park:18}. Especially, practical ergonomic design does not allow enough bandwidth of oblique projection angle and space in the temple side, which makes the glasses form factor even more stretched goal. 

\subsection*{Holographic displays using waveguide} 
There have been previous efforts to display holographic images through the light guiding slab for better form factor \cite{Yeom:2015}. Further efforts have been made to compensate aberration and improve image quality \cite{Yeom:2021, Lin:2018, Lin:2020}. However, there has been a fundamental limitation on scalability because the method is not intended to support pupil replication. The light guiding slab needs to be thick enough to avoid the replication, or the overlapped wavefront is scrambled, creating severe artifacts such as multiple ghost images and low contrast. Consequently, eyebox and field of view are limited to be small in such architectures.

In this study, we present a novel lens-less near-eye display system dubbed \textit{waveguide holography} that combines the merit of pupil-replicating waveguide and holographic display. Our approach is fundamentally different from previous works in that pupil expansion is enabled to provide large eyebox with about a millimeter thick waveguide. The core idea of waveguide holography is to model the coherent light interaction inside the pupil replicating waveguide. We introduce a novel multi-channel kernel modeling of waveguide wave propagation. The precise model calibration is enabled by complex wavefront capturing system and algorithm based on phase-shifting digital holography. As a result, the combination of the two state-of-art display technologies; waveguide image combiner and holographic displays, allows compact glasses like form factor, as well as displaying true 3D images. We demonstrate that out-coupled wavefront from the waveguide can be precisely controlled by modulating the input wavefront using our model. Our system is experimentally verified to support a full depth range with per-pixel depth reconstruction, diffraction limited resolution, as well as étendue expansion which enables a large 3D eyebox with a full native field of view of spatial light modulator with software-steered exit-pupil. We also present a detailed analysis and intuitive discussion for the architecture design and scalability. We conclude the study with a discussion on some limitations as well as interesting future works.

\section*{Results}

\subsection*{Architecture of waveguide holography}
Figure~\ref{fig:architecture}\textbf{a} illustrates the architecture of waveguide holography system, while Fig.~\ref{fig:architecture}\textbf{b} and \textbf{c} illustrate the compact prototype and benchtop prototype respectively. The system consists of a collimated laser light source, a spatial light modulator (SLM), polarizers, and a pupil replicating waveguide with surface relief gratings. Compared with the conventional waveguide display, the major difference is that the image projection module is replaced with the hologram projection module. Note that the SLM is placed without any projection lens, eliminating the need of physical propagation distance as well as achieving light weight. The benchtop prototype is built on the optical table with the same architecture and similar specifications, while the projection module is relayed with 4-$f$ imaging system. Benchtop prototype is useful for iterating design parameters and benchmarking the performance, while the compact prototype demonstrates its form factor. More details and further miniaturization methods are provided in the Method section.

The input light is modulated by the spatial light modulator and coupled by the in-coupler grating into the waveguide. The light propagates as total internal reflection mode and diffracted by exit-pupil expanding (EPE) grating and out-coupler grating that are typically designed as leaky gratings\cite{ayras2009exit}. This process generates manifold shifted copies of wavefront having different optical paths inside the waveguide, that interfere with each other so that the phase and intensity of the final output wavefront is intricately scrambled. In conventional waveguide image combiner, these phenomena are understood as coherence artifacts which should be avoided. However, we fully exploit this coherent interaction of light to precisely shape the output wavefront using spatial light modulator from the hologram projection module. 

Note that the étendue of transmitted light is expanded by the waveguide, but the bandwidth of information amount is unchanged. Therefore, shaping the entire output wavefront from modulating only input wavefront is fundamentally an over-constrained problem. To overcome the shortage of information bandwidth, we take advantage of the fact that most of the output wavefront does not enter the eye pupil. We set the virtual target aperture at the eyebox domain as a region of interest (ROI) for wavefront shaping, and this aperture can be computationally steered to match the size and 3D location of user’s eye pupil. This idea is similar with some of the previous CGH generation algorithms \cite{Georgiou:2008}. With the aid of eyetracking, the system can fully utilize the expanded étendue and achieve a software-steered eye-box without mechanical steering as large as conventional waveguide image combiners can provide.

\subsection*{Modeling of waveguide holography}

\begin{figure}[ht]
\centering
\includegraphics[width=\linewidth]{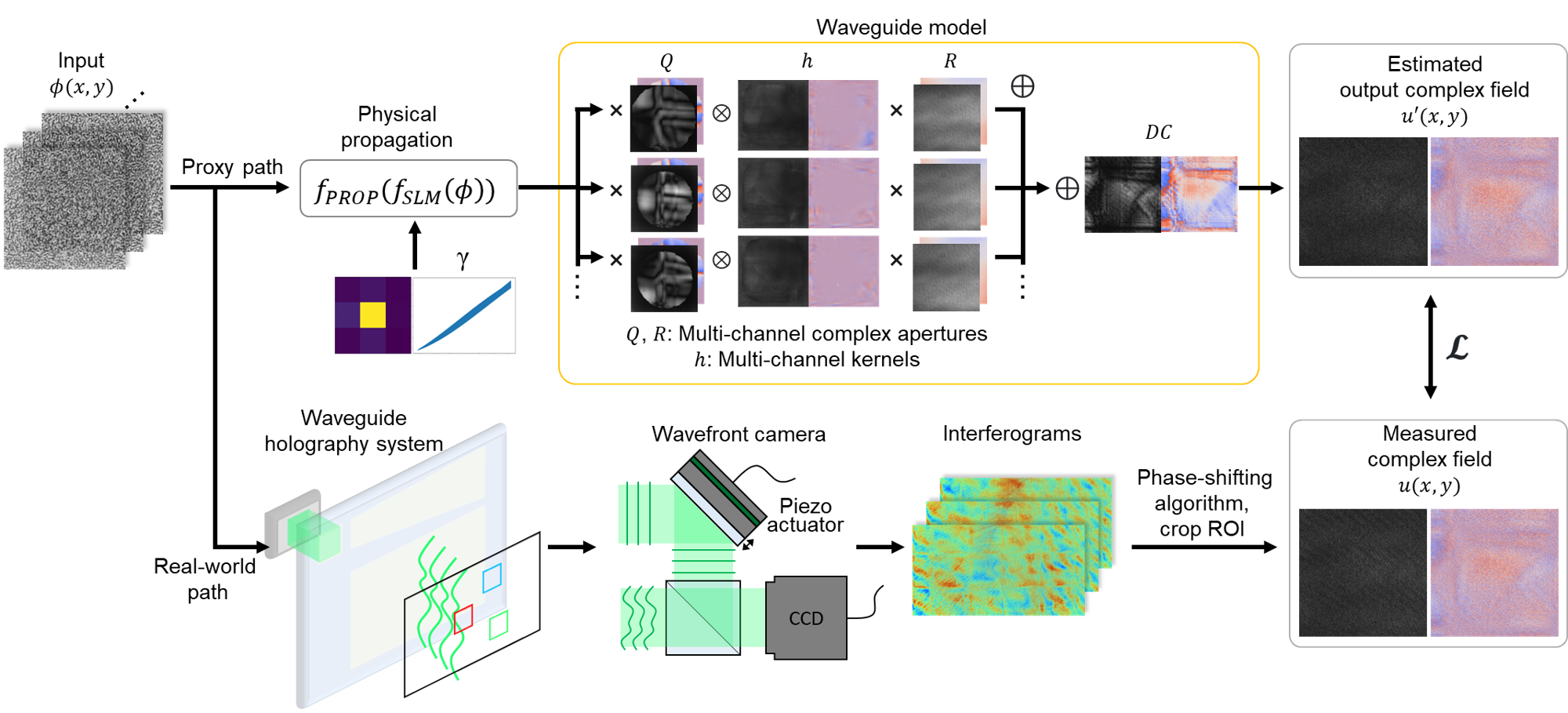}
\caption{Illustration of waveguide holography model pipeline. The upper part demonstrates the proxy path that models the waveguide propagation system, while the lower part demonstrates the real-world path that demonstrates the experimental pipeline to measure the out complex wavefront from the waveguide, which is used to generate the dataset for model calibration.}
\label{fig:pipeline}
\end{figure}
The core idea of waveguide holography is to model the coherent light interaction inside the pupil-replicating waveguide. We start with making a bold assumption that the waveguide can be approximated as a linear spatially invariant (LSI) system. The light in-coupling and out-coupling process of the waveguide can be simplified as combination of three major optical interactions; optical propagation in the waveguide substrate, total internal reflection at the substrate boundary, and the first order diffraction at the gratings. Note that all the three interactions are linear operators with complex-valued input and output. Also, spatially invariant property can be satisfied under the assumption of homogeneous grating profiles, in other words, each grating does not have optical power or boundary. Although there are physical boundaries of the gratings, this condition can be approximately satisfied to light paths that does not encounter grating boundaries.
LSI assumption enables key advantages to model the waveguide system in terms of wave optics regime. First, the whole complicated interactions can be simplified as a single convolution operation. This interpretation is computationally efficient compared to tracking all the light interactions of different paths inside the waveguide, which involves manifold operations with heavy computation. Based on the assumption, we build a differentiable forward model that is useful for model calibration and CGH rendering. The analytic derivation of the convolution kernel and its gradient is provided in the Supplementary Material.

Despite advantages of LSI assumption, the typical pupil replicating waveguides are not a perfect spatially invariant system in practical. There are plethora factors that alleviate the spatial invariant assumption, such as the non-uniformity of the grating, the surface flatness of the substrate, or the slant angle of the slab. Especially, physical boundaries of the grating introduce clipping of wavefront and edge diffraction, resulting different optical paths depending on the position at the input domain. In addition, defects in the grating and unwanted scattering at particles or dust all contribute to invalidating LSI approximation. 
Therefore, we introduce a multi-channel convolution model with complex apertures to handle the spatial-variant nature of the system. Our model pipeline is illustrated in the upper part of Fig.~\ref{fig:pipeline}, which consists of the multi-channel kernels $h$ and the complex apertures $Q$, $R$. All the apertures and kernels are complex valued 2D matrices and their sizes are dependent on the input SLM size and output ROI size. Aperture $Q$ is intended to model the in-coupler of the waveguide, and also helps to select different convolution path depending on the position at the in-coupler. Each $h$ is intended to emulate different possible light interaction paths inside the waveguide, which is the main source of spatial variance characteristics. Aperture $R$ additionally calibrates the intensity and phase fluctuation of the resultant field after the convolution, possibly caused by out-coupler grating or EPE grating. By merging output complex wavefronts from all the paths as a linear complex-number summation, the model acquires capacity to capture spatial-variant property. Also, linear summation induces smooth transition in between adjacent position to prohibit the model becoming too sensitive, while still maintaining the differentiable property. 
We also add the parameters to model SLM response and physical propagation of wavefront in front of the waveguide model. The SLM modeling consists of crosstalk kernel and spatially varying phase response function. The physical propagation includes free space propagation, as well as 3D tilt and homography change from the alignment mismatch and aberration. Details are provided in the Method/Supplementary Material.

\begin{figure}[ht]
\centering
\includegraphics[width=\linewidth]{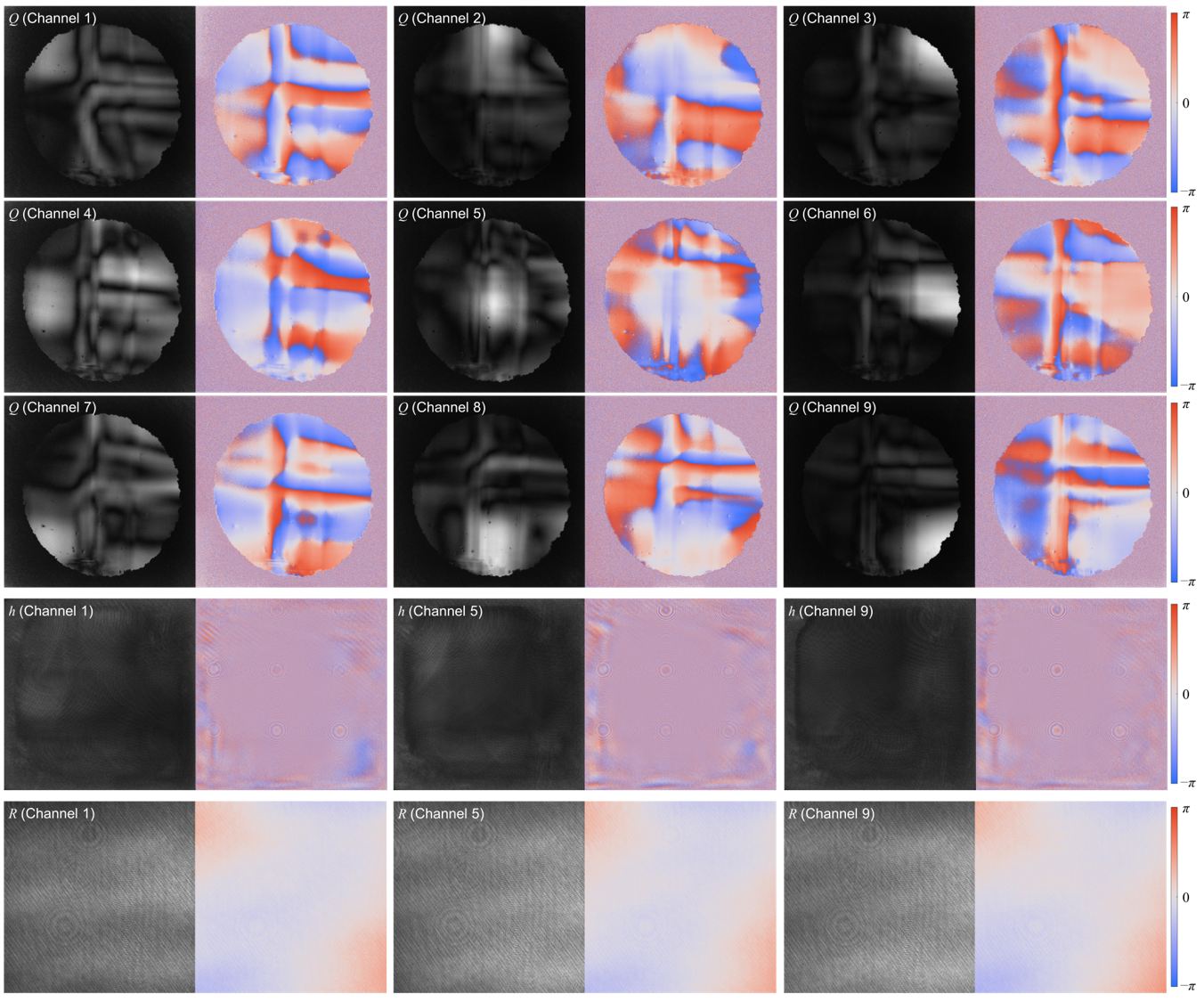}
\caption{Illustrations of optimized model parameters as gray-scale amplitude and pseudo-color map phase images: multi-channel complex apertures $Q$ (top), multi-channel kernels $h$ (middle), and $R$ (bottom). }
\label{fig:kernels}
\end{figure}

\begin{figure}[ht]
\centering
\includegraphics[width=\linewidth]{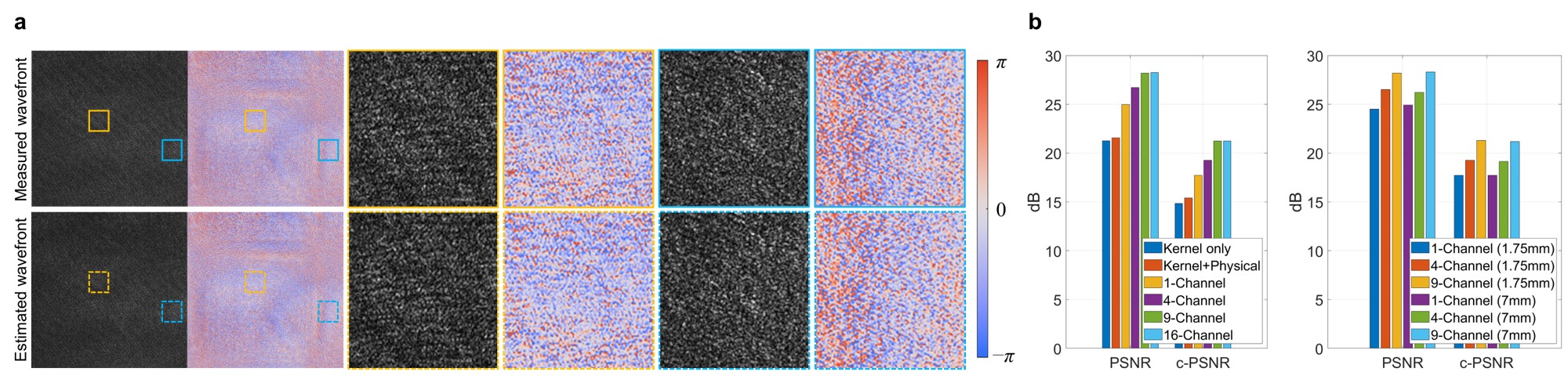}
\caption{\textbf{a} Comparison of the measured output wavefront and the estimated wavefront using the model. Orange and red insets are sampled at the same position in the eyebox domain, so as blue and green insets. Magnified images of insets are displayed in the right side for visual comparison. \textbf{b} The ablation analysis result of complex wavefront estimation performance according to the variations of the modeling (left) and scaling the size of the model, by changing the ROI size and the number of channel (right).}
\label{fig:estimated+ablation}
\end{figure}

\subsection*{Model calibration using complex wavefront camera}
The bottom part of Fig.~\ref{fig:pipeline} illustrates the complex wavefront dataset acquisition pipeline. We implement the Mach-Zehnder type phase-shifting interferometer system at the out-coupler side of the waveguide that captures the interferogram of the output wavefront from the waveguide and the plane reference beam \cite{MyungKim, Kim2011}. Then the phase-shifting algorithm is applied to retrieve the complex wavefront. We dubbed the interferometer system as wavefront camera for convenience. Random phase input is used for generating the dataset since it contains all the frequency components uniformly. After the data acquisition is finished, the loss is calculated as an L2 norm between estimated complex field and the measured complex field dataset during the training stage as:
\begin{equation}
\mathcal{L}=\|u^{'}(x,y)-u(x,y)\|^{2}.
\label{eq:loss}
\end{equation}
 
We emphasize that the complex wavefront capture is one of the key factors that enables the precise training of waveguide holography model. Compared with generic free space propagation, the light propagation inside waveguide results complicated overlapping and coherent interference that scrambles the amplitude and phase. By only measuring the amplitude, it is difficult to infer the waveguide kernels and complex apertures in the model. With wavefront camera, the access to phase information successfully retrieves the coherent light interaction in the waveguide.
Also, our method offers one-time calibration for large 3D eyebox area. Once the model is trained, pupil size, location, and position can be freely selected within the ROI by cropping different area from the estimated wavefront. In addition, eye-relief of the eyebox can be changed by numerically propagating the wavefront. This is a major difference from conventional camera-in-the-loop calibration methods which were impractical to calibrate all the possible pupil locations and size independently. The size of the ROI of the model can be selected depending on the wavefront camera’s sensor size and computationally practical model size. The larger ROI increases the usable 3D eyebox area after the one-time calibration, but it requires the larger memory to fit the model and the larger computation load. Up to 7 mm square eyebox has been fitted to the model, mainly restricted by the sensor size of the wavefront camera.

\subsection*{Ablation study}
Figure~\ref{fig:estimated+ablation} illustrates the result of output wavefront estimation. To evaluate the effectiveness of the different elements consisting the model, ablation analysis is performed as presented in the left of Fig.~\ref{fig:estimated+ablation}\textbf{b}, where the ROI is set as 3.5 mm square. First, the kernel-only model consists of only a single $h$ kernel while physical propagation module, $Q$, $R$ and $DC$ components are omitted. Then we add physical propagation module to the kernel-only model. The single channel indicates the full pipeline including $h$, $Q$, $R$ and $DC$ components as shown in Fig.~\ref{fig:pipeline}\textbf. We use peak signal-to-noise ratio (PSNR) and complex PSNR (c-PSNR) values to evaluate the similarity between estimated wavefront and the measured wavefront. The complex PSNR is calculated by concatenating real and imaginary part of the complex wavefront to form a real-valued matrix. Higher value indicates that the model predicts the output wavefront with a higher precision. The c-PSNR tends to be more sensitive than PSNR as only a slight change in the phase offset will result in a large error distance in the complex number domain. The result shows that having complex apertures and DC component is helpful to increase the fidelity of the model. Also, it can be verified that multi-channel model significantly boosts the performance compared with single-channel model. The effectiveness is eventually saturated as 9-channel and 16-channel do not show noticeable difference. 

In the right of Fig.~\ref{fig:estimated+ablation}\textbf{b}, we show the scalability of the model by varying the size of the output ROI. The size of the ROI does not significantly affect the model performance which aligns with the assumptions used in the modeling. At the in-coupler side, each channel that shares similar convolution path can be assorted spatially as shown in the shape of $Q$. Then the pupil replication process extends the receptive field of each channel to entire out-coupler area. Therefore the required number of channel is not dependent on the out-coupler domain, but dominantly decided at the in-coupler domain.


\subsection*{CGH rendering}
Once the model training is finished, the CGH can be calculated by adding a numerical propagation at the end of the model pipeline with parameterized input phase of the SLM. The input phase is initialized as random and propagates through forward path to generate the output retinal image. The loss is calculated as an L2 norm of the difference of the target image and the model output. For 3D contents, the loss can be calculated at multiple depths and added together. Focal stacks or light field images can be used to promote the accurate blurring effect\cite{Choi:2021, Lee:2022,kavakli2022realistic}. The loss is back propagated to update the input phase and the whole process is iterated until the estimated result reaches a certain PSNR value.

\subsection*{Experimental results}
\begin{figure}[ht]
\centering
\includegraphics[width=\linewidth]{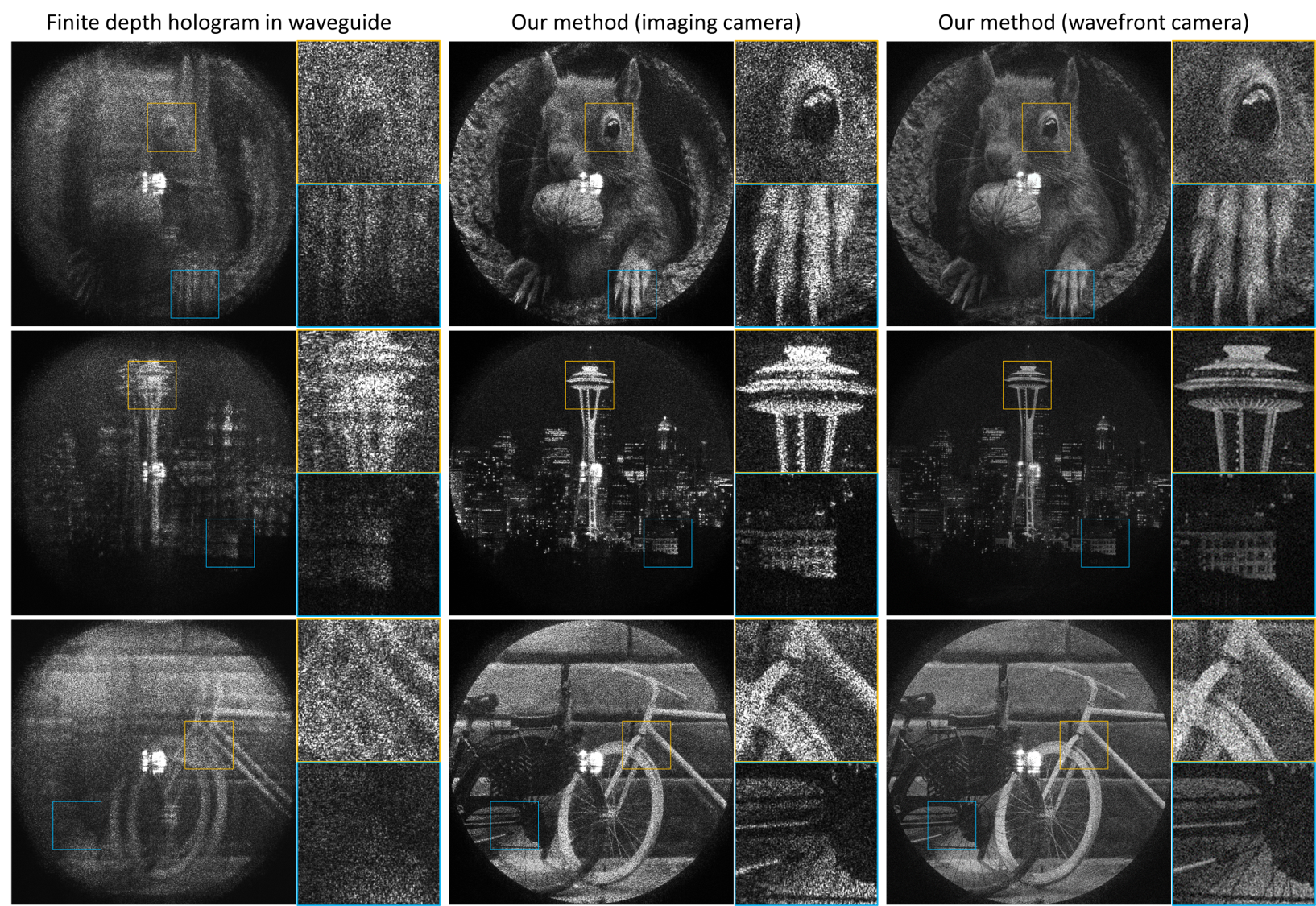}
\caption{Experimental results of waveguide holography. First column illustrates the artifact when projecting finite depth hologram in the waveguide without using our method. Second and third columns illustrate the result with hologram generated using our waveguide holography model, captured with imaging camera and wavefront camera respectively. The insets correspond to 1.3 degree of field of view and all the images are displayed at 3 diopter (D) from the user's pupil. Limited fill factor of the SLM generates a DC noise at the center of the field of view, which is discussed later. See Supplementary Material for more results. }
\label{fig:result}
\end{figure}

Figure~\ref{fig:result} demonstrates the experimental results of the waveguide holography. The images are captured in a benchtop prototype, and the field of view is slightly less than 8 degrees horizontally and vertically, decided by the SLM pitch size. Further system details are provided in the Method sections and Supplementary Material. The capturing is performed using two different methods. First, we put an imaging camera with the 3D printed entrance pupil mask with exact size and position of the targeted ROI and capture the image directly. Second, we capture the complex wavefront in the eyebox domain using a wavefront camera and numerically propagate to the image plane. Wavefront camera can avoid aberration from camera lens or alignment error because numerical propagation replaces physical aperture and camera lens. Also, it offers much larger depth range close to infinity thanks to its numerical refocusing capability. However, phase shifting  process could add noise to the reconstructed image. 
We use both methods to evaluate the results. The first column of Figure~\ref{fig:result} shows the artifact when the finite depth hologram is displayed via the waveguide for comparison. To render the CGH used in the first column, we replaced the waveguide module in the pipeline with generic wave propagation function to display image at the targeted depth without consideration of the waveguide. As expected, the images suffer severely from ghost noise and aberration created by duplicated pupils. The second and third column of Figure~\ref{fig:result} show the display results using our model captured with imaging camera and wavefront camera respectively. It can be clearly verified that the model successfully reconstruct the holograms at finite depth through the waveguide. Wavefront camera provides higher resolution with slightly reduced contrast as explained above.

\begin{figure}[h!]
\centering
\includegraphics[width=\linewidth]{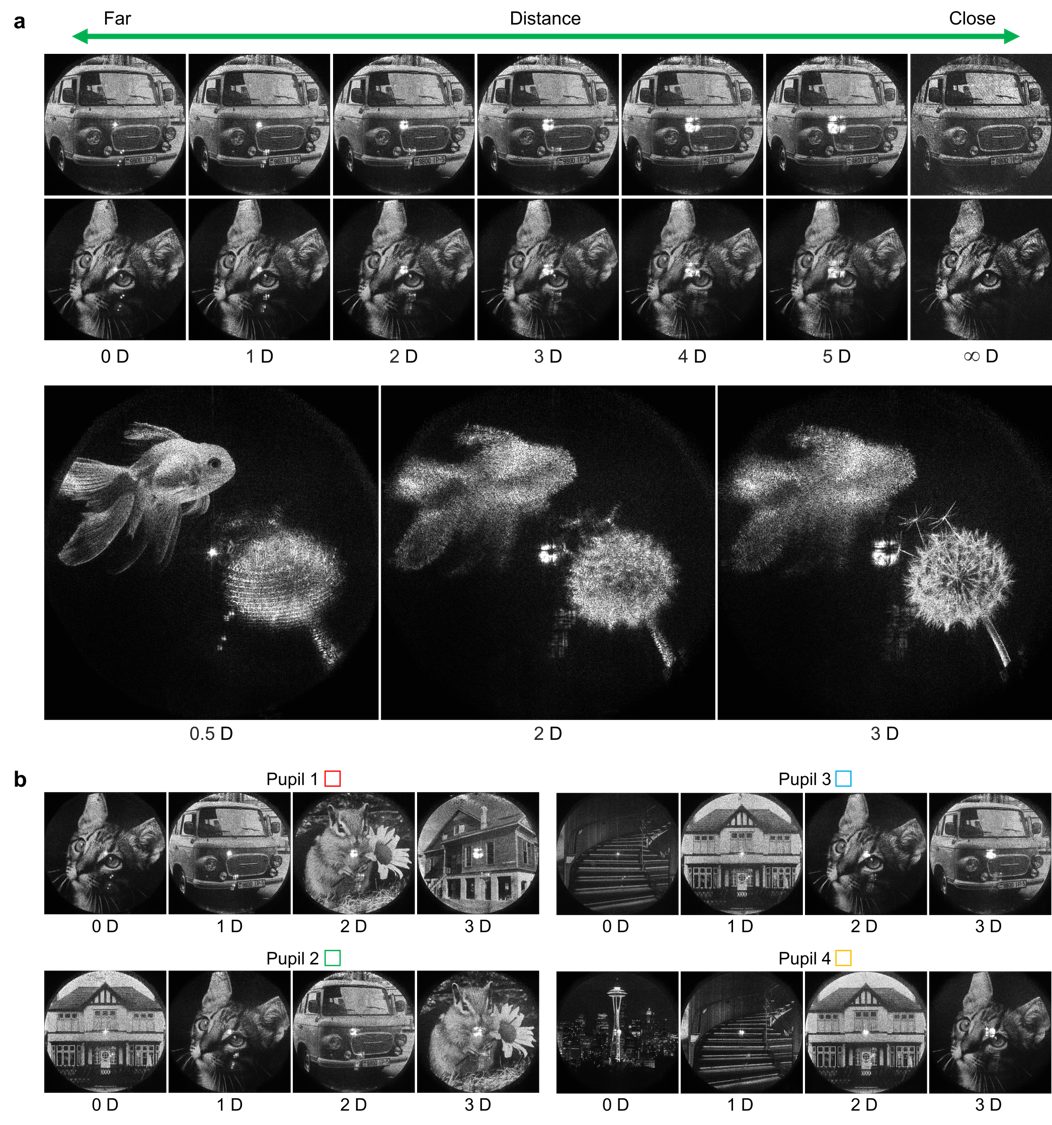}
\caption{Display results that demonstrate the depth range and eyebox range. \textbf{a} Holograms are displayed at different depths from 0 diopter to infinity diopter. The infinity diopter indicates the pupil plane and the image is captured directly putting the camera sensor without lens. All three fish and dandelion images are displayed with a single 3D hologram and captured at the different depths. \textbf{b} Results captured at different pupil locations within the large eyebox to demonstrate étendue expansion. The holograms are optimized for corresponding pupil locations. Their positions and more results are provided in the Supplementary Material.}
\label{fig:result2}
\end{figure}

\subsubsection*{Depth range and 3D eyebox}
We demonstrate that our prototype can generate image at arbitrary depth from zero to infinity distance from the eyebox plane, confirming the full 3D image can be displayed in the waveguide only modulating the SLM. Figure~\ref{fig:result2}\textbf{a} shows the display results at different depths to verify the depth range of the system. It is noteworthy that our model improves the image quality significantly even when the image is displayed at the infinity depth, where there is no explicit presence of ghost noise (see Supplementary Material). 

Figure~\ref{fig:result2}\textbf{b} demonstrates the large eyebox of the system by forming software-steered eyebox at arbitrarily sampled pupil locations. Once the model is calibrated, any size and location in 3D space of the eyebox can be chosen without additional pupil calibration. Each pupil successfully displays images at different depths. The lens-less hologram projection configuration provides good uniformity over eyebox domain. We present related simulation in the following section. 

\subsubsection*{Compact prototype result}

\begin{figure}[ht]
\centering
\includegraphics[width=300pt]{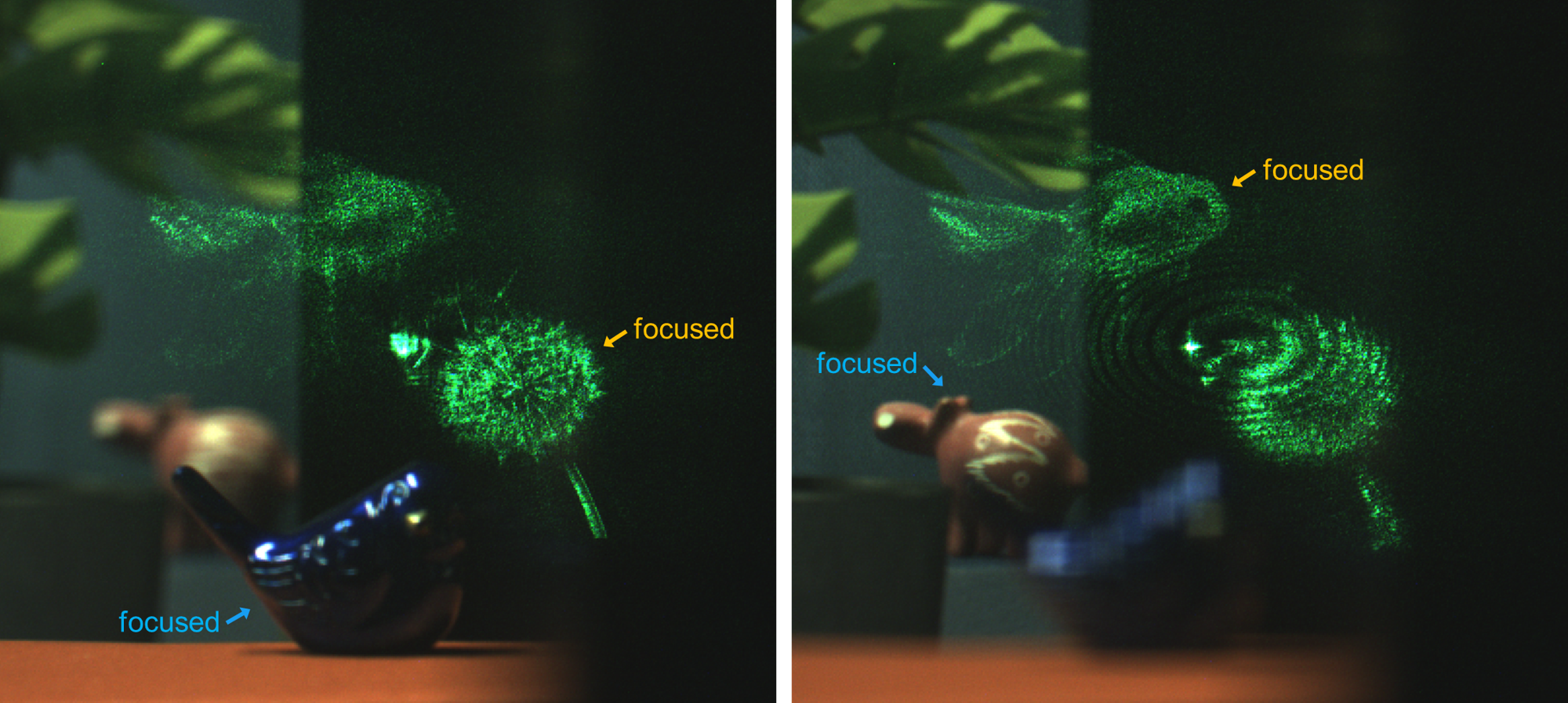}
\caption{Augmented reality display results captured through the compact prototype glasses. Good see-through quality can be achieved with $>90\%$ transmittance. The dandelion image is displayed at 3 diopter and the fish is displayed at 1 diopter.  }
\label{fig:ar}
\end{figure}
Display results of the compact prototype are presented in Fig.~\ref{fig:ar}. The scene is captured through the waveguide to demonstrate the see-through quality. The image is more noisy compared to the benchtop prototype because calibration fidelity is degraded due to SLM artifacts. Details are discussed in the Method section.

\subsubsection*{Overcoming the resolution limit of the waveguide display}
In theory, the angular resolution of the waveguide display is decided by the number of the modes and mode spacing that a waveguide can support for the monochromatic light as: 
\begin{equation}\label{res}
\begin{aligned}
\delta\theta_{res} = \frac{\lambda}{2 t  \tan{\theta_{T}}},
\end{aligned}
\end{equation}
where $t$ is the thickness of the substrate and $\theta_{T}$ is TIR angle of the FOV component. However, this does not hold in typical waveguide displays because a lot of non-idealities in the system, such as aberration from the projection module or surface flatness. 
Besides, numerous beam clippings at the edges of the gratings during the replication process as well as clipping at the user's eye pupil reduce the effective numerical aperture of the output beam. This beam clipping effect has been an inevitable degradation factor that sets the fundamental limit of the resolution in the waveguide display system in the most cases.

\begin{figure}[ht]
\centering
\includegraphics[width=\linewidth]{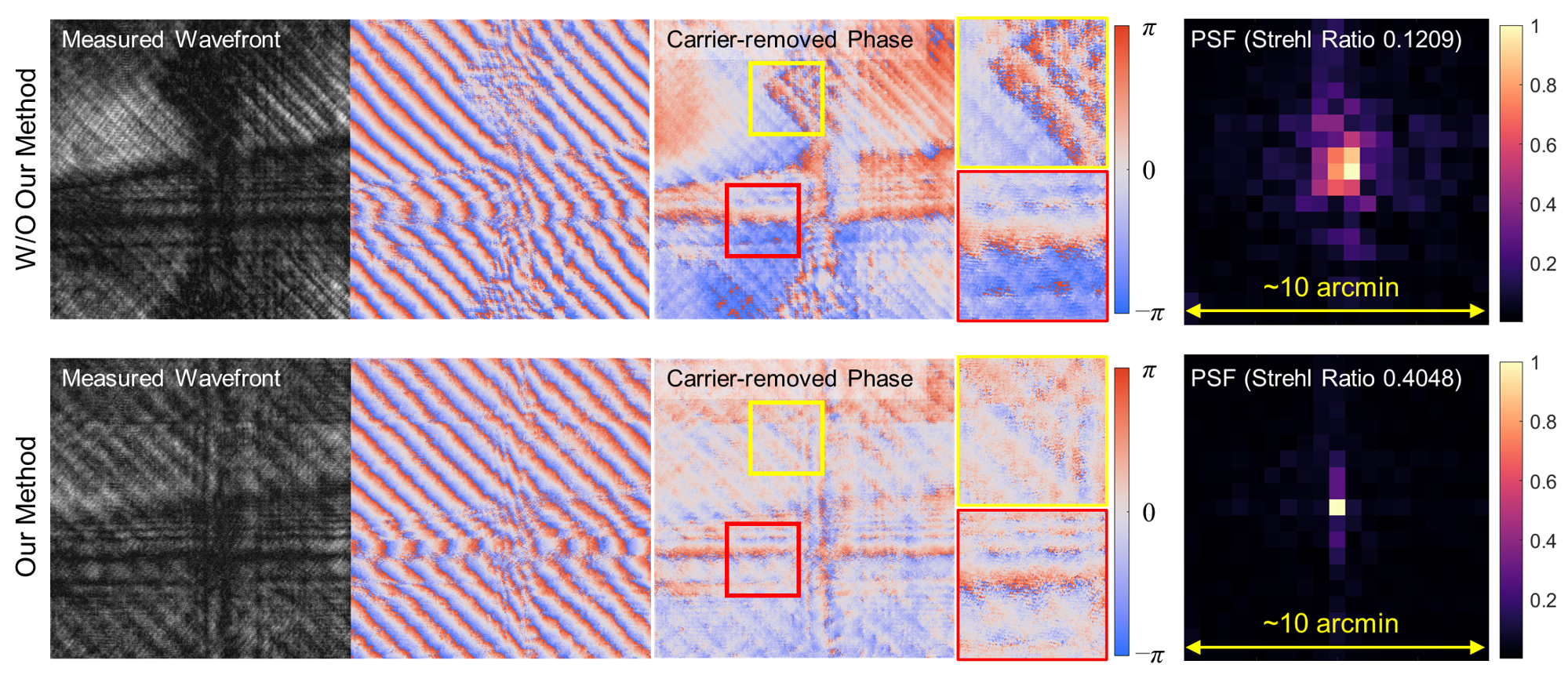}
\caption{Illustration of resolution enhancement capability of waveguide holography. The upper row and bottom row show the experimentally captured output wavefront using a tilted plane wave as a target which forms a single image point at the infinity distance, without and with using our method, respectively. In the upper row, the target tilted plane wave is used as an input wavefront to the waveguide without the knowledge of the waveguide model. In the bottom row, the input wavefront is optimized by the complex loss function to form the target tilted plane wave using our model. In the middle, the carrier frequency is removed for the visualization of phase discontinuities by dividing with the target wavefront phase. In the right, the PSFs are obtained by applying Fourier transform to the measured wavefronts, where a single pixel is equivalent to $0.53$ arcmin.}
\label{fig:super}
\end{figure}

We demonstrate that our method can overcome such resolution limit. With the knowledge of light interaction in the waveguide, the phase discontinuities caused by beam clippings can be stitched to achieve smooth phase in the eyebox. Figure.~\ref{fig:super} illustrates the experimental results to display a tilted plane wave target, captured by wavefront camera. 
Without using our method, the phase discontinuities with similar shapes of grating boundaries are observed in the wavefront. With the optimization, it can be visually verified that the phase discontinuity is minimized over the pupil. Besides, the amplitude is optimized to be more uniform, as the destructive interference is forced to be minimized to generate uniform plane beam output. In the right of Fig.~\ref{fig:super}, the point spread function (PSF) is illustrated which clearly visualizes the improvement of the resolution. As a result, sub-arc-minute resolution is achieved with over threefold increased Strehl ratio. Strehl ratio is calculated using Mahajan formula \cite{Mahajan:1982} on the carrier-removed phase. The results suggest that our method opens up the new possibilities of fully utilizing coherent nature of light. 

\subsection*{Design space and scalability analysis}

The design of our waveguide shares general goals of conventional pupil replicating waveguides, such as high image resolution and high light throughput efficiency, as well as the uniformity of out-coupled light intensity in both eyebox domain and field of view domain. Among the various parameters involved in the design space, we focus on pupil replication distance $d_{rep}$ which is a function of thickness $t$ and the TIR angle $\theta_{TIR}$ of corresponding field of view component as:
\begin{equation}\label{eq1}
d_{rep} =2 t \tan{\theta_{TIR}}.
\end{equation}
If $d_{rep}$ is set too large, some field of view component may not fill the eye pupil and cause vignetting artifact or partial loss of the image. Otherwise, when the pupil replication distance is too smaller than the beam size, the intensity of guided light decays too fast because it requires too many TIRs for the light propagation over the entire eyebox. 
In the conventional waveguide displays, the resolution tends to be degraded with thinner waveguide as effective numerical aperture is reduced and more clipping happens. In waveguide holography, the numerical aperture is not necessarily degraded as demonstrated in the previous section. 
However, it imposes additional restriction in the wavefront optimization. If the wavefront is replicated and overlapped too densely, it reduces the degree of freedom to control the interference in a desired manner, which trades the image quality. 

\begin{figure}[h!]
\centering
\includegraphics[width=\linewidth]{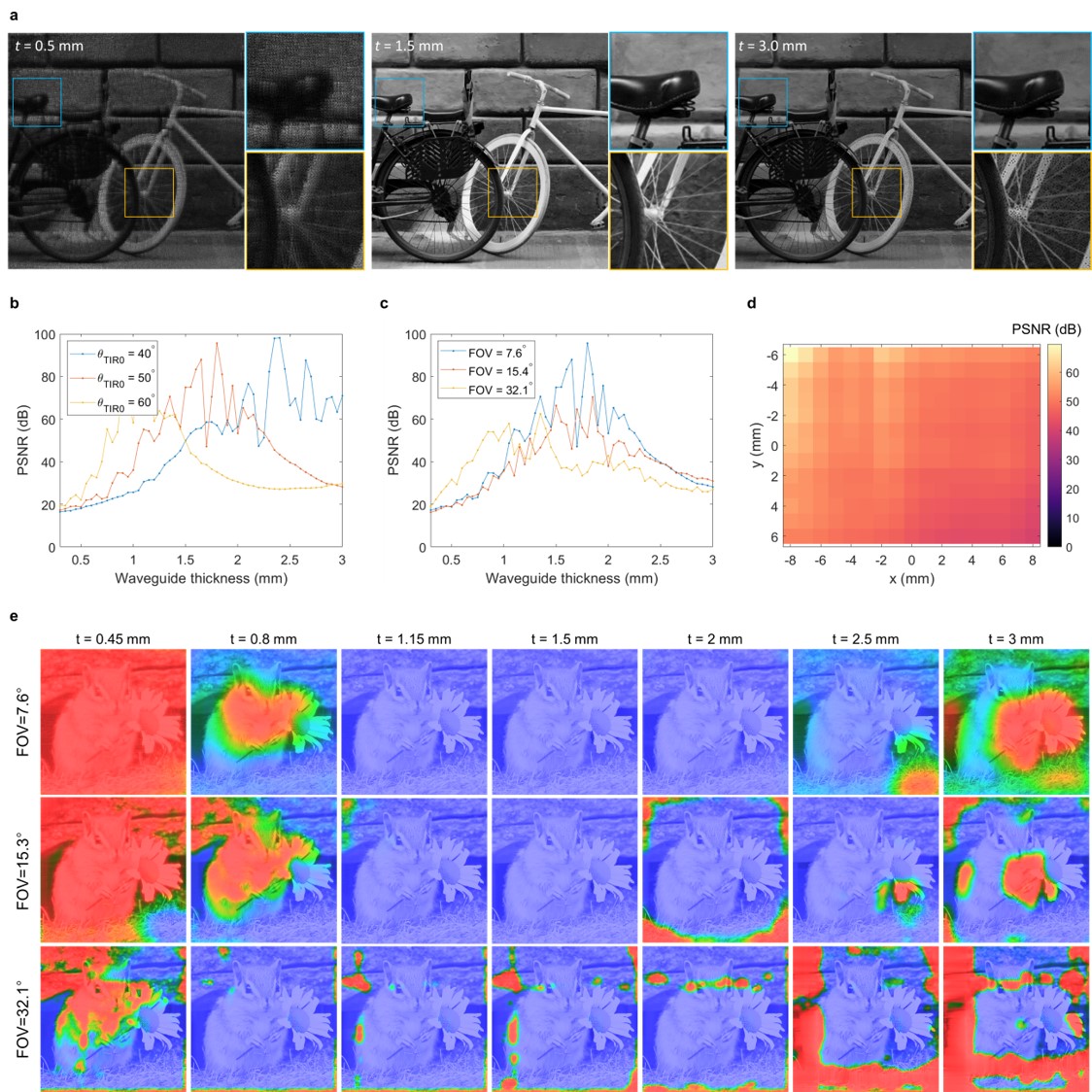}
\caption{\textbf{a} Simulated display results with different waveguide thickness, while other system parameters remain identical as the benchtop prototype. The artifacts are demonstrated when the waveguide thickness is too thin or too thick. \textbf{b} A plot of simulated display performance sweeping the waveguide thickness, with three different TIR propagation angles of the center field of view $\theta_{TIR0}$. The display performance is evaluated as an average PSNR value of five different images simulated to be displayed at 1 m. For 40 degrees of center propagation angle, refractive index is adjusted to 1.7 because 1.5 refractive index cannot support full field of view as total internal reflection mode. \textbf{c} Simulated display performance plot varying the field of view to evaluate the scalability of the system. The pixel pitch is set as 4 µm, 2 µm, and 1 µm respectively, from the smallest FOV. For 1 µm pixel pitch, the refractive index of waveguide is set as 2 to support the full field of view. \textbf{d} Visualization of the eyebox uniformity simulation in the benchtop prototype. \textbf{e} Visualization of field of view uniformity according to varying waveguide thickness and FOV, using pseudo colormap of hdrvdp2 metric\cite{hdrvdp2}. The red area indicates that the observer would perceive the artifact with a high probability compared to the original image.  }
\label{fig:thickness}
\end{figure}

Both artifacts can be observed in Figure.~\ref{fig:thickness}\textbf{a}, which illustrates the simulated display results with different waveguide thickness using the specifications of the benchtop prototype. 
The parameter sweep results of the display performance varying the thickness and the center TIR propagation angle are plotted in Fig.~\ref{fig:thickness}\textbf{b}. The plot suggests that there is a range of sweet spot that achieves the best image quality, where $d_{rep}$ is balanced between pupil density and the wavefront optimization freedom. 
Figure.~\ref{fig:thickness}\textbf{c} demonstrates the scalability of the system in terms of field of view varying the pixel pitch of the SLM. As supported field of view grows, the maximum PSNR tends to drop slightly.
The results Fig.~\ref{fig:thickness}\textbf{a}--\textbf{c} are simulated in a center eyebox location. The uniformity of the image quality in the eyebox domain is visualized in Fig.~\ref{fig:thickness}\textbf{d} and over 40 dB of PSNR value is predicted in the large eyebox area as designed. Also, we present the visualization of the field of view uniformity in Fig.~\ref{fig:thickness}\textbf{e}. When the waveguide thickness is too thin, the artifact tends to be dependent on the texture of displayed image since the cause is limited optimization freedom. While the thickness is too thick, the artifact tends to show arbitrary pattern independent to the contents because the vignetting from the pupil density is the main reason. The simulation results offer a valuable intuition for optimizing the design parameters of the system, as well as showing the scalability of the architecture. 


\section*{Discussion}
In this study, we propose a novel AR near-eye display architecture by combining two state-of-art display technologies; waveguide display, which is the industry norm technology aiming augmented reality glasses, and holographic display, which is believed to be the ultimate 3D display technology. We have demonstrated to display holographic images at arbitrary depths and eyebox position through pupil replicating waveguide for the first time, opening the new possibility towards the true 3D holographic glasses. Here we discuss existing limitations and interesting challenges for future research topics.

In our design, the FOV is limited by the pixel-pitch of the SLM because we use a direct view hologram projection configuration without projection lens. There are ongoing efforts in the academia and industry to develop and manufacture sub-micrometer pixel pitch panels\cite{kim2018development, hwang202021} with growing expectation of AR/VR, which will be the long term solution for the FOV. Although, we note that there are alternative variations of hologram projection module to increase FOV immediately. For example, a projection lens can be used similarly to a conventional image projection module, so the FOV is limited by the focal length of the lens and the size of the SLM. 
Nevertheless, we choose to demonstrate the feasibility of the ultimate lens-free architecture, betting on the future breakthrough in the micro display technology.  
Currently, the limited fill factor of the SLM creates DC noise peak at the center of the FOV. This is not easy to eliminate completely by incremental enhancement of the fill factor since all the DC energy are focused into the center at the infinity conjugate. An alternative solution is to offset the projection angle of the SLM and filter the DC noise with a thick volume grating \cite{Bang:2019}. Depending on the waveguide design, the center FOV may be re-arranged to the periphery to fully utilize the bandwidth of the SLM \cite{mukawa20088}. 
A full color implementation can be achieved with the waveguide with RGB channels and utilizing temporal multiplexing.
Additionally, we note that calibration process is sensitive to the mechanical perturbation caused by vibration or airflow by the nature of interferometer system. We empirically observed that the system is relatively more robust in the display stage than the calibration data acquisition stage. A further investigation on the mechanical sensitivity of the system will be useful. 
Lastly, collimated light source is supplied from external system in our prototypes. For further miniaturization, a single mode waveguide or a beam expanding wedge prism could generate a collimated light source in a compact form factor \cite{Xiong:2013}. 

There are several interesting future works to be explored. First, more precise modeling of the waveguide system can be studied. Our model is built on the physical intuition of the waveguide propagation process, consisting of interpretable model parameters without a black box-like component of the model or a redundant calibration process. Such approach allows useful performance analysis and optimization of the design which will be helpful for understanding the scalability and system requirements. We concluded that waveguide holography shares very similar design rules as conventional waveguide, however imposing additional restriction for wavefront optimization. We expect further refined physically accurate model will not only improve the display quality, but also facilitate system optimization and co-design of the waveguide. Or, more efficient modeling is another direction to be pursued which will be beneficial for the real-time display. We find there are some redundancies in the model parameters, as such each kernel shares similarities in amplitude and phase shapes and the gain from increasing number of channel saturates at some point. Such redundancies can be reduced to shrink down the size and computation load of the model.
Additionally, the system can take advantage of advanced features of SLM such as high framerate or complex modulation capability\cite{complexmodulation}. Several studies have recently reported that utilization of temporal multiplexing helps speckle reduction and image quality enhancement \cite{Lee:2022,choi2022time}. Such method can be utilized to improve the image quality. Besides, it is expected by a simulation that the display performance increases a lot when the SLM has complex modulation capability.

We finish the discussion by suggesting further applications of the work. We expect the methodologies used in the work can be extended to various purposes. The complex wavefront measurement method can be a useful tool for calibrating other complex display systems using the coherent light source. 
Also, the multi-channel model can be modified to capture other aspects of light interaction, for example, modeling of partial coherence modes. Lastly, our work demonstrates a potential of laser light source for conventional pupil replicating waveguide display. Display industry is currently making a lot of effort to develop bright and efficient light sources for the waveguide display projection module. Laser light source can satisfy both qualities, however it has not been used in practice because of the coherence artifact which is difficult to control. In our experiments, it was demonstrated that not only coherent artifacts can be mitigated, but also that resolution limitations can be overcome by fully exploiting coherent nature of light. We hope our contributions would facilitate more follow-up research to take further steps towards the ultra-compact, true 3D holographic AR glasses.

\section*{Methods}

 \begin{figure}[ht]
\centering
\includegraphics[width=\linewidth]{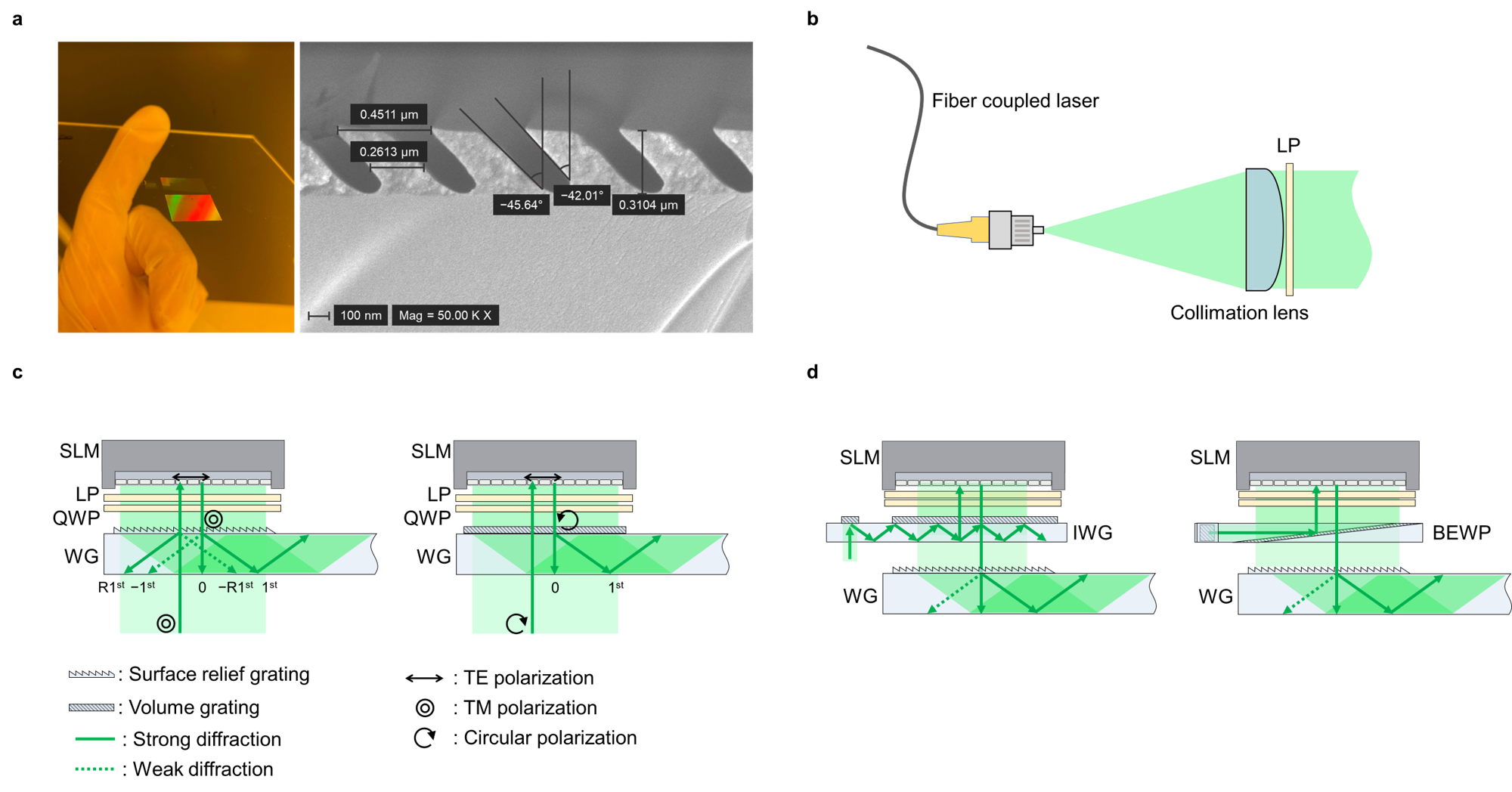}
\caption{Illustrations of further details of the system and potential miniaturization strategies. \textbf{a} Photographs of waveguide sample used in the prototypes. \textbf{b} Beam expanding system for plane wave illumination. \textbf{c} Illustrations of hologram projection module without the beam splitter, with respect to type of in-coupler grating. Left figure shows surface relief grating while right figure shows polarization volume grating which does not generate unwanted diffraction. \textbf{d} Example designs of miniaturized illumination module. Left figure illustrates illumination waveguide (IWG) and the right figure illustrates beam expanding wedge prism (BEWP)\cite{Xiong:2013}.       }
\label{fig:waveguide}
\end{figure}
\subsection*{Waveguide fabrication}
The waveguide is fabricated with glass substrate with refractive index of 1.5 and 1.15 mm thickness. The thickness of the substrate is selected based on the simulation results presented in Fig.~\ref{fig:thickness}\textbf{b} to achieve a good performance and still retain the thin form factor. The center TIR angle is set as 50 degrees with the center wavelength of 532nm wavelength. The waveguide is designed to support 20 degrees of horizontal field of view and 16 $\times$ 12 mm of eyebox size. The waveguide samples are fabricated using nano-imprinting method which is suitable for mass production. The specifications of surface relief gratings such as shape, slant angle, and aspect ratio are fine-tuned using rigorous simulation to achieve spatial and angular uniformity at the eyebox domain. In general, targeting higher uniformity reduces the grating efficiency and thus trades overall efficiency. We set the merit function to balance between uniformity and efficiency, to achieve over 5\% of end-to-end throughput efficiency in average and maximize the uniformity. The photographs of fabricated waveguide sample and measured specifications of the grating structure are presented in Fig.~\ref{fig:waveguide}\textbf{a}. The grating structure is designed as a saw-tooth shape to minimize the unwanted diffraction orders\cite{tamir1977analysis, wang2004compact}, however it diffracts some portion of light as $-1^{st}$ order. Therefore we used a beam splitter in the hologram projection module and the light source is expanded outside of the system as shown in Fig.~\ref{fig:waveguide}\textbf{b}. If the input beam is illuminated through the waveguide without the beam splitter, the reverse path at the in-coupler will generate  unwanted diffraction term that is mixed with the signal as shown in the left of Fig.~\ref{fig:waveguide}\textbf{c}, indicated as $-R1^{st}$. 
For further miniaturization, using volume grating in-couplers such as Bragg volume grating or polarization volume grating can eliminate unwanted diffraction orders due to high selectivity as shown in the right of Fig.~\ref{fig:waveguide}\textbf{c}. Lastly, Fig.~\ref{fig:waveguide}\textbf{d} illustrates the miniaturization methods of illumination module as described in the discussion. 

\subsection*{Prototype implementation}
In the optical benchtop prototype, we used 532 nm Cobolt Samba 1500 mW laser as a light source, Piezosystem Jena PZ-38 as a piezo actuator for phase shifting digital holography, and Meadowlark E-series 1920 × 1200 SLM. In de-magnifying relay system, 150mm (L1) and 75mm (L2) focal length lenses are used. 62.5\% the SLM area (1200 $\times$ 1200 pixels) is used to generate input wavefront. We built two wavefront cameras in the system; one for capturing waveguide output wavefront, and the other for capturing relayed SLM to optimize SLM parameters and homography in the model. Both wavefront cameras share the same piezo actuator using a beam splitter for phase shifting. A neutral density (ND) filter is used to attenuate the light intensity in the reference path for both wavefront cameras. To capture the result images, a 3D printed pupil aperture and an imaging camera are mounted on 3-axis motorized stages and placed at the copy of the output wavefront duplicated using a beam splitter. Optical benchtop prototype has 7.64 degrees of field of view in both horizontal and vertical directions, 16 $\times$ 12 mm eyebox size with 3D eyebox.

In the compact prototype, de-magnifying 4-$f$ relay system is eliminated and instead we used 4K SLM with 3840 $\times$ 2160 resolution and 3.74 µm pixel pitch, supporting 8.2 degrees of field of view. Only 20\% of the SLM area (1300 $\times$ 1300 pixels) is actively used while other pixels are deactivated. The image quality degradation in the compact prototype is majorly caused by the SLM performance. The SLM has about 10\% of phase flicker, which severely degrades the fidelity of model calibration compared with the benchtop prototype since the complex wavefront calibration method is highly sensitive to the phase error. Poorly calibrated $DC$ component causes interference pattern artifact at the far distance. Also, the SLM has more severe fringe field effect that further deteriorates the image quality. We identified that un-filtered high-order diffraction from the SLM is not a major cause of the quality degradation. 
The compact prototype experiment is performed on the optical table and collimated laser is provided externally through a 5 mm beam splitter to the hologram projection module. For calibration, another beam splitter is placed at the eyebox of the waveguide to combine the reference beam. 

\subsection*{Details of the algorithm}
The physical propagation module in the Fig.~\ref{fig:pipeline} consists of crosstalk kernel of the adjacent SLM pixels, spatially varying phase modulation function of the SLM ($\gamma$), free space propagation with 3D tilt~\cite{Matsushima:09,Matsushima:03}, and homography transformation function. The crosstalk kernel has $3\times3$ size and $\gamma$ is modeled as a polynomial with 18 coefficients. Free space propagation function has 3 parameters including 2D tilt angles and the propagation distance. The homography transformation function is up to second orders with 12 coefficients. The physical propagation module is calibrated in advance in the benchtop prototype using wavefront camera placed at the relayed SLM and then used as the initial value of the model calibration to accelerate the calibration. 
In the waveguide model, the size of $Q$ is selected to cover the physical size of the in-coupler grating, which is 1200$\times$1200. The size of $R$ and $DC$ is selected to be equal to the ROI size. The side length of Kernel $h$ is set as summation of $Q$ and $R$. 
For the model calibration, about a thousand captured wavefronts are used as a dataset. In the CGH rendering stage, the loss converged around 1000 iterations and we ran up to 3000 iterations. 
Further implementation details are provided in the Supplementary Material.

\bibliography{sample}

\begin{thebibliography}{10}
\urlstyle{rm}
\expandafter\ifx\csname url\endcsname\relax
  \def\url#1{\texttt{#1}}\fi
\expandafter\ifx\csname urlprefix\endcsname\relax\def\urlprefix{URL }\fi
\expandafter\ifx\csname doiprefix\endcsname\relax\def\doiprefix{DOI: }\fi
\providecommand{\bibinfo}[2]{#2}
\providecommand{\eprint}[2][]{\url{#2}}

\bibitem{Azuma1997}
\bibinfo{author}{Azuma, R.~T.}
\newblock \bibinfo{journal}{\bibinfo{title}{A survey of augmented reality}}.
\newblock {\emph{\JournalTitle{Presence: Teleoperators and virtual
  environments}}} \textbf{\bibinfo{volume}{6}}, \bibinfo{pages}{335--385}
  (\bibinfo{year}{1997}).

\bibitem{Kress2013}
\bibinfo{author}{Kress, B.} \& \bibinfo{author}{Starner, T.}
\newblock \bibinfo{journal}{\bibinfo{title}{A review of head-mounted displays
  (hmd) technologies and applications for consumer electronics}}.
\newblock {\emph{\JournalTitle{In Proceedings of SPIE Defense, Security, and
  Sensing, International Society for Optics and Photonics}}}
  (\bibinfo{year}{2013}).

\bibitem{bleereview}
\bibinfo{author}{Lee, B.}, \bibinfo{author}{Jo, Y.}, \bibinfo{author}{Yoo, D.}
  \& \bibinfo{author}{Lee, J.}
\newblock \bibinfo{title}{{Recent progresses of near-eye display for AR and
  VR}}.
\newblock In \bibinfo{editor}{Stella, E.} (ed.)
  \emph{\bibinfo{booktitle}{Multimodal Sensing and Artificial Intelligence:
  Technologies and Applications II}}, vol. \bibinfo{volume}{11785},
  \bibinfo{pages}{1178503}, \doiprefix\url{10.1117/12.2596128}.
  \bibinfo{organization}{International Society for Optics and Photonics}
  (\bibinfo{publisher}{SPIE}, \bibinfo{year}{2021}).

\bibitem{levola20067}
\bibinfo{author}{Levola, T.}
\newblock \bibinfo{title}{7.1: Invited paper: Novel diffractive optical
  components for near to eye displays}.
\newblock In \emph{\bibinfo{booktitle}{SID Symposium Digest of Technical
  Papers}}, vol.~\bibinfo{volume}{37}, \bibinfo{pages}{64--67}
  (\bibinfo{organization}{Wiley Online Library}, \bibinfo{year}{2006}).

\bibitem{amitai2010substrate}
\bibinfo{author}{Amitai, Y.}
\newblock \bibinfo{title}{Substrate-guided optical devices}
  (\bibinfo{year}{2010}).
\newblock \bibinfo{note}{US Patent 7,672,055}.

\bibitem{cheng2014design}
\bibinfo{author}{Cheng, D.}, \bibinfo{author}{Wang, Y.}, \bibinfo{author}{Xu,
  C.}, \bibinfo{author}{Song, W.} \& \bibinfo{author}{Jin, G.}
\newblock \bibinfo{journal}{\bibinfo{title}{Design of an ultra-thin near-eye
  display with geometrical waveguide and freeform optics}}.
\newblock {\emph{\JournalTitle{Optics express}}} \textbf{\bibinfo{volume}{22}},
  \bibinfo{pages}{20705--20719} (\bibinfo{year}{2014}).

\bibitem{xu2019methods}
\bibinfo{author}{Xu, M.} \& \bibinfo{author}{Hua, H.}
\newblock \bibinfo{journal}{\bibinfo{title}{Methods of optimizing and
  evaluating geometrical lightguides with microstructure mirrors for augmented
  reality displays}}.
\newblock {\emph{\JournalTitle{Optics express}}} \textbf{\bibinfo{volume}{27}},
  \bibinfo{pages}{5523--5543} (\bibinfo{year}{2019}).

\bibitem{kress201711}
\bibinfo{author}{Kress, B.~C.} \& \bibinfo{author}{Cummings, W.~J.}
\newblock \bibinfo{title}{11-1: invited paper: towards the ultimate mixed
  reality experience: Hololens display architecture choices}.
\newblock In \emph{\bibinfo{booktitle}{SID symposium digest of technical
  papers}}, vol.~\bibinfo{volume}{48}, \bibinfo{pages}{127--131}
  (\bibinfo{organization}{Wiley Online Library}, \bibinfo{year}{2017}).

\bibitem{Gu_pvhwaveguide}
\bibinfo{author}{Gu, Y.} \emph{et~al.}
\newblock \bibinfo{journal}{\bibinfo{title}{Holographic waveguide display with
  large field of view and high light efficiency based on polarized volume
  holographic grating}}.
\newblock {\emph{\JournalTitle{IEEE Photonics Journal}}}
  \textbf{\bibinfo{volume}{14}}, \bibinfo{pages}{1--7},
  \doiprefix\url{10.1109/JPHOT.2021.3127547} (\bibinfo{year}{2022}).

\bibitem{ayras2009exit}
\bibinfo{author}{{\"A}yr{\"a}s, P.}, \bibinfo{author}{Saarikko, P.} \&
  \bibinfo{author}{Levola, T.}
\newblock \bibinfo{journal}{\bibinfo{title}{Exit pupil expander with a large
  field of view based on diffractive optics}}.
\newblock {\emph{\JournalTitle{Journal of the Society for Information
  Display}}} \textbf{\bibinfo{volume}{17}}, \bibinfo{pages}{659--664}
  (\bibinfo{year}{2009}).

\bibitem{kress2020optical}
\bibinfo{author}{Kress, B.~C.}
\newblock \bibinfo{title}{Optical architectures for augmented-, virtual-, and
  mixed-reality headsets} (\bibinfo{organization}{Society of Photo-Optical
  Instrumentation Engineers}, \bibinfo{year}{2020}).

\bibitem{erickson2020exploring}
\bibinfo{author}{Erickson, A.}, \bibinfo{author}{Kim, K.},
  \bibinfo{author}{Bruder, G.} \& \bibinfo{author}{Welch, G.~F.}
\newblock \bibinfo{title}{Exploring the limitations of environment lighting on
  optical see-through head-mounted displays}.
\newblock In \emph{\bibinfo{booktitle}{Symposium on Spatial User Interaction}},
  \bibinfo{pages}{1--8} (\bibinfo{year}{2020}).

\bibitem{chang2020toward}
\bibinfo{author}{Chang, C.}, \bibinfo{author}{Bang, K.},
  \bibinfo{author}{Wetzstein, G.}, \bibinfo{author}{Lee, B.} \&
  \bibinfo{author}{Gao, L.}
\newblock \bibinfo{journal}{\bibinfo{title}{Toward the next-generation vr/ar
  optics: a review of holographic near-eye displays from a human-centric
  perspective}}.
\newblock {\emph{\JournalTitle{Optica}}} \textbf{\bibinfo{volume}{7}},
  \bibinfo{pages}{1563--1578} (\bibinfo{year}{2020}).

\bibitem{shi2022design}
\bibinfo{author}{Shi, X.} \emph{et~al.}
\newblock \bibinfo{journal}{\bibinfo{title}{Design of a dual focal-plane
  near-eye display using diffractive waveguides and multiple lenses}}.
\newblock {\emph{\JournalTitle{Applied Optics}}} \textbf{\bibinfo{volume}{61}},
  \bibinfo{pages}{5844--5849} (\bibinfo{year}{2022}).

\bibitem{Yoo:19}
\bibinfo{author}{Yoo, C.} \emph{et~al.}
\newblock \bibinfo{journal}{\bibinfo{title}{Dual-focal waveguide see-through
  near-eye display with polarization-dependent lenses}}.
\newblock {\emph{\JournalTitle{Opt. Lett.}}} \textbf{\bibinfo{volume}{44}},
  \bibinfo{pages}{1920--1923}, \doiprefix\url{10.1364/OL.44.001920}
  (\bibinfo{year}{2019}).

\bibitem{nam2020aberration}
\bibinfo{author}{Nam, S.-W.} \emph{et~al.}
\newblock \bibinfo{journal}{\bibinfo{title}{Aberration-corrected full-color
  holographic augmented reality near-eye display using a pancharatnam-berry
  phase lens}}.
\newblock {\emph{\JournalTitle{Optics Express}}} \textbf{\bibinfo{volume}{28}},
  \bibinfo{pages}{30836--30850} (\bibinfo{year}{2020}).

\bibitem{kim2021vision}
\bibinfo{author}{Kim, D.} \emph{et~al.}
\newblock \bibinfo{journal}{\bibinfo{title}{Vision-correcting holographic
  display: evaluation of aberration correcting hologram}}.
\newblock {\emph{\JournalTitle{Biomedical Optics Express}}}
  \textbf{\bibinfo{volume}{12}}, \bibinfo{pages}{5179--5195}
  (\bibinfo{year}{2021}).

\bibitem{Kavakli:21}
\bibinfo{author}{Kavakl{i}, K.} \emph{et~al.}
\newblock \bibinfo{journal}{\bibinfo{title}{Pupil steering holographic display
  for pre-operative vision screening of cataracts}}.
\newblock {\emph{\JournalTitle{Biomed. Opt. Express}}}
  \textbf{\bibinfo{volume}{12}}, \bibinfo{pages}{7752--7764},
  \doiprefix\url{10.1364/BOE.439545} (\bibinfo{year}{2021}).

\bibitem{Maimone:2017}
\bibinfo{author}{Maimone, A.}, \bibinfo{author}{Georgiou, A.} \&
  \bibinfo{author}{Kollin, J.~S.}
\newblock \bibinfo{journal}{\bibinfo{title}{Holographic near-eye displays for
  virtual and augmented reality}}.
\newblock {\emph{\JournalTitle{ACM Trans. Graph.}}}
  \textbf{\bibinfo{volume}{36}}, \doiprefix\url{10.1145/3072959.3073624}
  (\bibinfo{year}{2017}).

\bibitem{Shi:2021}
\bibinfo{author}{Shi, L.}, \bibinfo{author}{Li, B.}, \bibinfo{author}{Kim, C.},
  \bibinfo{author}{Kellnhofer, P.} \& \bibinfo{author}{Matusik, W.}
\newblock \bibinfo{journal}{\bibinfo{title}{Towards real-time photorealistic 3d
  holography with deep neural networks}}.
\newblock {\emph{\JournalTitle{Nature}}} \textbf{\bibinfo{volume}{592}}
  (\bibinfo{year}{2021}).

\bibitem{Choi:2021}
\bibinfo{author}{Choi, S.}, \bibinfo{author}{Gopakumar, M.},
  \bibinfo{author}{Peng, Y.}, \bibinfo{author}{Kim, J.} \&
  \bibinfo{author}{Wetzstein, G.}
\newblock \bibinfo{journal}{\bibinfo{title}{Neural 3d holography: Learning
  accurate wave propagation models for 3d holographic virtual and augmented
  reality displays}}.
\newblock {\emph{\JournalTitle{ACM Trans. Graph.}}}
  \textbf{\bibinfo{volume}{40}}, \doiprefix\url{10.1145/3478513.3480542}
  (\bibinfo{year}{2021}).

\bibitem{Peng:2020:NeuralHolography}
\bibinfo{author}{Peng, Y.}, \bibinfo{author}{Choi, S.},
  \bibinfo{author}{Padmanaban, N.} \& \bibinfo{author}{Wetzstein, G.}
\newblock \bibinfo{journal}{\bibinfo{title}{{Neural Holography with
  Camera-in-the-loop Training}}}.
\newblock {\emph{\JournalTitle{ACM Trans. Graph. (SIGGRAPH Asia)}}}
  (\bibinfo{year}{2020}).

\bibitem{aksit2022perceptually}
\bibinfo{author}{Akşit, K.} \emph{et~al.}
\newblock \bibinfo{title}{Perceptually guided computer-generated holography}.
\newblock In \emph{\bibinfo{booktitle}{Advances in Display Technologies XII}},
  vol. \bibinfo{volume}{12024}, \bibinfo{pages}{11--14}
  (\bibinfo{organization}{SPIE}, \bibinfo{year}{2022}).

\bibitem{gracekuo}
\bibinfo{author}{Kuo, G.}, \bibinfo{author}{Waller, L.}, \bibinfo{author}{Ng,
  R.} \& \bibinfo{author}{Maimone, A.}
\newblock \bibinfo{journal}{\bibinfo{title}{High resolution \'{E}tendue
  expansion for holographic displays}}.
\newblock {\emph{\JournalTitle{ACM Trans. Graph.}}}
  \textbf{\bibinfo{volume}{39}}, \doiprefix\url{10.1145/3386569.3392414}
  (\bibinfo{year}{2020}).

\bibitem{chakravarthula2019wirtinger}
\bibinfo{author}{Chakravarthula, P.}, \bibinfo{author}{Peng, Y.},
  \bibinfo{author}{Kollin, J.}, \bibinfo{author}{Fuchs, H.} \&
  \bibinfo{author}{Heide, F.}
\newblock \bibinfo{journal}{\bibinfo{title}{Wirtinger holography for near-eye
  displays}}.
\newblock {\emph{\JournalTitle{ACM Transactions on Graphics (TOG)}}}
  \textbf{\bibinfo{volume}{38}}, \bibinfo{pages}{213} (\bibinfo{year}{2019}).

\bibitem{jpegpleno:21}
\bibinfo{author}{Muhamad, R.~K.} \emph{et~al.}
\newblock \bibinfo{journal}{\bibinfo{title}{Jpeg pleno holography: scope and
  technology validation procedures}}.
\newblock {\emph{\JournalTitle{Appl. Opt.}}} \textbf{\bibinfo{volume}{60}},
  \bibinfo{pages}{641--651}, \doiprefix\url{10.1364/AO.404305}
  (\bibinfo{year}{2021}).

\bibitem{yang2022perceptually}
\bibinfo{author}{Yang, F.} \emph{et~al.}
\newblock \bibinfo{journal}{\bibinfo{title}{Perceptually motivated loss
  functions for computer generated holographic displays}}.
\newblock {\emph{\JournalTitle{Scientific reports}}}
  \textbf{\bibinfo{volume}{12}}, \bibinfo{pages}{1--12} (\bibinfo{year}{2022}).

\bibitem{park2019ultrathin}
\bibinfo{author}{Park, J.}, \bibinfo{author}{Lee, K.} \& \bibinfo{author}{Park,
  Y.}
\newblock \bibinfo{journal}{\bibinfo{title}{Ultrathin wide-angle large-area
  digital 3d holographic display using a non-periodic photon sieve}}.
\newblock {\emph{\JournalTitle{Nature Communications}}}
  \textbf{\bibinfo{volume}{10}}, \bibinfo{pages}{1--8} (\bibinfo{year}{2019}).

\bibitem{samsung20}
\bibinfo{author}{An, J.} \emph{et~al.}
\newblock \bibinfo{journal}{\bibinfo{title}{Slim-panel holographic video
  display}}.
\newblock {\emph{\JournalTitle{Nature Communications}}}
  \textbf{\bibinfo{volume}{11}}, \bibinfo{pages}{5568},
  \doiprefix\url{10.1038/s41467-020-19298-4} (\bibinfo{year}{2020}).

\bibitem{peng2021partiallycoherent}
\bibinfo{author}{Peng, Y.}, \bibinfo{author}{Choi, S.}, , \bibinfo{author}{Kim,
  J.} \& \bibinfo{author}{Wetzstein, G.}
\newblock \bibinfo{journal}{\bibinfo{title}{Speckle-free holography with
  partially coherent light sources and camera-in-the-loop calibration}}.
\newblock {\emph{\JournalTitle{Science Advances}}}  (\bibinfo{year}{2021}).

\bibitem{kim2022holographicglasses}
\bibinfo{author}{Kim, J.} \emph{et~al.}
\newblock \bibinfo{title}{Holographic glasses for virtual reality}.
\newblock In \emph{\bibinfo{booktitle}{Proceedings of the ACM SIGGRAPH}},
  \bibinfo{pages}{1–8} (\bibinfo{year}{2022}).

\bibitem{Matsushima:05}
\bibinfo{author}{Matsushima, K.}
\newblock \bibinfo{journal}{\bibinfo{title}{Computer-generated holograms for
  three-dimensional surface objects with shade and texture}}.
\newblock {\emph{\JournalTitle{Appl. Opt.}}} \textbf{\bibinfo{volume}{44}},
  \bibinfo{pages}{4607--4614}, \doiprefix\url{10.1364/AO.44.004607}
  (\bibinfo{year}{2005}).

\bibitem{Kaczorowski:16}
\bibinfo{author}{Kaczorowski, A.}, \bibinfo{author}{Gordon, G. S.~D.} \&
  \bibinfo{author}{Wilkinson, T.~D.}
\newblock \bibinfo{journal}{\bibinfo{title}{Adaptive, spatially-varying
  aberration correction for real-time holographic projectors}}.
\newblock {\emph{\JournalTitle{Opt. Express}}} \textbf{\bibinfo{volume}{24}},
  \bibinfo{pages}{15742--15756}, \doiprefix\url{10.1364/OE.24.015742}
  (\bibinfo{year}{2016}).

\bibitem{chen2015improved}
\bibinfo{author}{Chen, J.-S.} \& \bibinfo{author}{Chu, D.}
\newblock \bibinfo{journal}{\bibinfo{title}{Improved layer-based method for
  rapid hologram generation and real-time interactive holographic display
  applications}}.
\newblock {\emph{\JournalTitle{Optics express}}} \textbf{\bibinfo{volume}{23}},
  \bibinfo{pages}{18143--18155} (\bibinfo{year}{2015}).

\bibitem{shimobaba2015review}
\bibinfo{author}{Shimobaba, T.}, \bibinfo{author}{Kakue, T.} \&
  \bibinfo{author}{Ito, T.}
\newblock \bibinfo{journal}{\bibinfo{title}{Review of fast algorithms and
  hardware implementations on computer holography}}.
\newblock {\emph{\JournalTitle{IEEE Transactions on Industrial Informatics}}}
  \textbf{\bibinfo{volume}{12}}, \bibinfo{pages}{1611--1622}
  (\bibinfo{year}{2015}).

\bibitem{lee2020deep}
\bibinfo{author}{Lee, J.} \emph{et~al.}
\newblock \bibinfo{journal}{\bibinfo{title}{Deep neural network for multi-depth
  hologram generation and its training strategy}}.
\newblock {\emph{\JournalTitle{Optics Express}}} \textbf{\bibinfo{volume}{28}},
  \bibinfo{pages}{27137--27154} (\bibinfo{year}{2020}).

\bibitem{mengu2016non}
\bibinfo{author}{Mengu, D.}, \bibinfo{author}{Ulusoy, E.} \&
  \bibinfo{author}{Urey, H.}
\newblock \bibinfo{journal}{\bibinfo{title}{Non-iterative phase hologram
  computation for low speckle holographic image projection}}.
\newblock {\emph{\JournalTitle{Optics Express}}} \textbf{\bibinfo{volume}{24}},
  \bibinfo{pages}{4462--4476} (\bibinfo{year}{2016}).

\bibitem{Kozacki:12}
\bibinfo{author}{Kozacki, T.}, \bibinfo{author}{Finke, G.},
  \bibinfo{author}{Garbat, P.}, \bibinfo{author}{Zaperty, W.} \&
  \bibinfo{author}{Kujawi\'{n}ska, M.}
\newblock \bibinfo{journal}{\bibinfo{title}{Wide angle holographic display
  system with spatiotemporal multiplexing}}.
\newblock {\emph{\JournalTitle{Opt. Express}}} \textbf{\bibinfo{volume}{20}},
  \bibinfo{pages}{27473--27481}, \doiprefix\url{10.1364/OE.20.027473}
  (\bibinfo{year}{2012}).

\bibitem{Jang:2018}
\bibinfo{author}{Jang, C.}, \bibinfo{author}{Bang, K.}, \bibinfo{author}{Li,
  G.} \& \bibinfo{author}{Lee, B.}
\newblock \bibinfo{journal}{\bibinfo{title}{Holographic near-eye display with
  expanded eye-box}}.
\newblock {\emph{\JournalTitle{ACM Trans. Graph.}}}
  \textbf{\bibinfo{volume}{37}}, \doiprefix\url{10.1145/3272127.3275069}
  (\bibinfo{year}{2018}).

\bibitem{Park:18}
\bibinfo{author}{Park, J.-H.} \& \bibinfo{author}{Kim, S.-B.}
\newblock \bibinfo{journal}{\bibinfo{title}{Optical see-through holographic
  near-eye-display with eyebox steering and depth of field control}}.
\newblock {\emph{\JournalTitle{Opt. Express}}} \textbf{\bibinfo{volume}{26}},
  \bibinfo{pages}{27076--27088}, \doiprefix\url{10.1364/OE.26.027076}
  (\bibinfo{year}{2018}).

\bibitem{Yeom:2015}
\bibinfo{author}{Yeom, H.-J.} \emph{et~al.}
\newblock \bibinfo{journal}{\bibinfo{title}{3d holographic head mounted display
  using holographic optical elements with astigmatism aberration
  compensation}}.
\newblock {\emph{\JournalTitle{Opt. Express}}} \textbf{\bibinfo{volume}{23}},
  \bibinfo{pages}{32025--32034}, \doiprefix\url{10.1364/OE.23.032025}
  (\bibinfo{year}{2015}).

\bibitem{Yeom:2021}
\bibinfo{author}{Yeom, J.}, \bibinfo{author}{Son, Y.} \& \bibinfo{author}{Choi,
  K.}
\newblock \bibinfo{journal}{\bibinfo{title}{Crosstalk reduction in voxels for a
  see-through holographic waveguide by using integral imaging with compensated
  elemental images}}.
\newblock {\emph{\JournalTitle{Photonics}}} \textbf{\bibinfo{volume}{8}},
  \doiprefix\url{10.3390/photonics8060217} (\bibinfo{year}{2021}).

\bibitem{Lin:2018}
\bibinfo{author}{Lin, W.-K.}, \bibinfo{author}{Matoba, O.},
  \bibinfo{author}{Lin, B.-S.} \& \bibinfo{author}{Su, W.-C.}
\newblock \bibinfo{journal}{\bibinfo{title}{Astigmatism and deformation
  correction for a holographic head-mounted display with a wedge-shaped
  holographic waveguide}}.
\newblock {\emph{\JournalTitle{Appl. Opt.}}} \textbf{\bibinfo{volume}{57}},
  \bibinfo{pages}{7094--7101}, \doiprefix\url{10.1364/AO.57.007094}
  (\bibinfo{year}{2018}).

\bibitem{Lin:2020}
\bibinfo{author}{Lin, W.-K.}, \bibinfo{author}{Matoba, O.},
  \bibinfo{author}{Lin, B.-S.} \& \bibinfo{author}{Su, W.-C.}
\newblock \bibinfo{journal}{\bibinfo{title}{Astigmatism correction and quality
  optimization of computer-generated holograms for holographic waveguide
  displays}}.
\newblock {\emph{\JournalTitle{Opt. Express}}} \textbf{\bibinfo{volume}{28}},
  \bibinfo{pages}{5519--5527}, \doiprefix\url{10.1364/OE.381193}
  (\bibinfo{year}{2020}).

\bibitem{Georgiou:2008}
\bibinfo{author}{Georgiou, A.}, \bibinfo{author}{Christmas, J.},
  \bibinfo{author}{Collings, N.}, \bibinfo{author}{Moore, J.} \&
  \bibinfo{author}{Crossland, W.}
\newblock \bibinfo{journal}{\bibinfo{title}{Aspects of hologram calculation for
  video frames}}.
\newblock {\emph{\JournalTitle{Journal of Optics A: Pure and Applied Optics}}}
  \textbf{\bibinfo{volume}{10}}, \bibinfo{pages}{035302},
  \doiprefix\url{10.1088/1464-4258/10/3/035302} (\bibinfo{year}{2008}).

\bibitem{MyungKim}
\bibinfo{author}{Kim, M.~K.}
\newblock \bibinfo{journal}{\bibinfo{title}{{Principles and techniques of
  digital holographic microscopy}}}.
\newblock {\emph{\JournalTitle{SPIE Reviews}}} \textbf{\bibinfo{volume}{1}},
  \bibinfo{pages}{018005}, \doiprefix\url{10.1117/6.0000006}
  (\bibinfo{year}{2010}).

\bibitem{Kim2011}
\bibinfo{author}{Kim, M.~K.}
\newblock \emph{\bibinfo{title}{Digital Holographic Microscopy}},
  \bibinfo{pages}{149--190} (\bibinfo{publisher}{Springer New York},
  \bibinfo{address}{New York, NY}, \bibinfo{year}{2011}).

\bibitem{Lee:2022}
\bibinfo{author}{Lee, B.}, \bibinfo{author}{Kim, D.}, \bibinfo{author}{Lee,
  S.}, \bibinfo{author}{Chen, C.} \& \bibinfo{author}{Lee, B.}
\newblock \bibinfo{journal}{\bibinfo{title}{High-contrast, speckle-free, true
  3d holography via binary cgh optimization}}.
\newblock {\emph{\JournalTitle{Scientific Reports}}}
  \textbf{\bibinfo{volume}{12}},
  \doiprefix\url{doi.org/10.1038/s41598-022-06405-2} (\bibinfo{year}{2022}).

\bibitem{kavakli2022realistic}
\bibinfo{author}{Kavakl{\i}, K.}, \bibinfo{author}{Itoh, Y.},
  \bibinfo{author}{Urey, H.} \& \bibinfo{author}{Akşit, K.}
\newblock \bibinfo{journal}{\bibinfo{title}{Realistic defocus blur for
  multiplane computer-generated holography}}.
\newblock {\emph{\JournalTitle{arXiv preprint arXiv:2205.07030}}}
  \doiprefix\url{https://doi.org/10.48550/arXiv.2205.07030}
  (\bibinfo{year}{2022}).

\bibitem{Mahajan:1982}
\bibinfo{author}{Mahajan, V.~N.}
\newblock \bibinfo{journal}{\bibinfo{title}{Strehl ratio for primary
  aberrations: some analytical results for circular and annular pupils}}.
\newblock {\emph{\JournalTitle{J. Opt. Soc. Am.}}}
  \textbf{\bibinfo{volume}{72}}, \bibinfo{pages}{1258--1266},
  \doiprefix\url{10.1364/JOSA.72.001258} (\bibinfo{year}{1982}).

\bibitem{hdrvdp2}
\bibinfo{author}{Mantiuk, R.}, \bibinfo{author}{Kim, K.~J.},
  \bibinfo{author}{Rempel, A.~G.} \& \bibinfo{author}{Heidrich, W.}
\newblock \bibinfo{journal}{\bibinfo{title}{Hdr-vdp-2: A calibrated visual
  metric for visibility and quality predictions in all luminance conditions}}.
\newblock {\emph{\JournalTitle{ACM Trans. Graph.}}}
  \textbf{\bibinfo{volume}{30}}, \doiprefix\url{10.1145/2010324.1964935}
  (\bibinfo{year}{2011}).

\bibitem{kim2018development}
\bibinfo{author}{Kim, Y.-H.} \emph{et~al.}
\newblock \bibinfo{journal}{\bibinfo{title}{Development of high-resolution
  active matrix spatial light modulator}}.
\newblock {\emph{\JournalTitle{Optical Engineering}}}
  \textbf{\bibinfo{volume}{57}}, \bibinfo{pages}{061606}
  (\bibinfo{year}{2018}).

\bibitem{hwang202021}
\bibinfo{author}{Hwang, C.-S.} \emph{et~al.}
\newblock \bibinfo{title}{21-2: Invited paper: 1$\mu$m pixel pitch spatial
  light modulator panel for digital holography}.
\newblock In \emph{\bibinfo{booktitle}{SID Symposium Digest of Technical
  Papers}}, vol.~\bibinfo{volume}{51}, \bibinfo{pages}{297--300}
  (\bibinfo{organization}{Wiley Online Library}, \bibinfo{year}{2020}).

\bibitem{Bang:2019}
\bibinfo{author}{Bang, K.}, \bibinfo{author}{Jang, C.} \& \bibinfo{author}{Lee,
  B.}
\newblock \bibinfo{journal}{\bibinfo{title}{Compact noise-filtering volume
  gratings for holographic displays}}.
\newblock {\emph{\JournalTitle{Opt. Lett.}}} \textbf{\bibinfo{volume}{44}},
  \bibinfo{pages}{2133--2136}, \doiprefix\url{10.1364/OL.44.002133}
  (\bibinfo{year}{2019}).

\bibitem{mukawa20088}
\bibinfo{author}{Mukawa, H.} \emph{et~al.}
\newblock \bibinfo{title}{8.4: distinguished paper: a full color eyewear
  display using holographic planar waveguides}.
\newblock In \emph{\bibinfo{booktitle}{SID Symposium Digest of Technical
  Papers}}, vol.~\bibinfo{volume}{39}, \bibinfo{pages}{89--92}
  (\bibinfo{organization}{Wiley Online Library}, \bibinfo{year}{2008}).

\bibitem{Xiong:2013}
\bibinfo{author}{Xiong, Y.} \emph{et~al.}
\newblock \bibinfo{journal}{\bibinfo{title}{Coherent backlight system for
  flat-panel holographic 3d display}}.
\newblock {\emph{\JournalTitle{Optics Communications}}}
  \textbf{\bibinfo{volume}{296}}, \bibinfo{pages}{41--46}
  (\bibinfo{year}{2013}).

\bibitem{complexmodulation}
\bibinfo{author}{Jang, S.-W.} \emph{et~al.}
\newblock \bibinfo{journal}{\bibinfo{title}{Complex spatial light modulation
  capability of a dual layer in-plane switching liquid crystal panel}}.
\newblock {\emph{\JournalTitle{Scientific Reports}}}
  \textbf{\bibinfo{volume}{12}},
  \doiprefix\url{https://doi.org/10.1038/s41598-022-12292-4}
  (\bibinfo{year}{2022}).

\bibitem{choi2022time}
\bibinfo{author}{Choi, S.} \emph{et~al.}
\newblock \bibinfo{title}{Time-multiplexed neural holography: A flexible
  framework for holographic near-eye displays with fast heavily-quantized
  spatial light modulators}.
\newblock In \emph{\bibinfo{booktitle}{Proceedings of the ACM SIGGRAPH}},
  \bibinfo{pages}{1–8} (\bibinfo{year}{2022}).

\bibitem{tamir1977analysis}
\bibinfo{author}{Tamir, T.} \& \bibinfo{author}{Peng, S.-T.}
\newblock \bibinfo{journal}{\bibinfo{title}{Analysis and design of grating
  couplers}}.
\newblock {\emph{\JournalTitle{Applied physics}}}
  \textbf{\bibinfo{volume}{14}}, \bibinfo{pages}{235--254}
  (\bibinfo{year}{1977}).

\bibitem{wang2004compact}
\bibinfo{author}{Wang, B.}, \bibinfo{author}{Jiang, J.} \&
  \bibinfo{author}{Nordin, G.~P.}
\newblock \bibinfo{journal}{\bibinfo{title}{Compact slanted grating couplers}}.
\newblock {\emph{\JournalTitle{Optics express}}} \textbf{\bibinfo{volume}{12}},
  \bibinfo{pages}{3313--3326} (\bibinfo{year}{2004}).

\bibitem{Matsushima:09}
\bibinfo{author}{Matsushima, K.} \& \bibinfo{author}{Shimobaba, T.}
\newblock \bibinfo{journal}{\bibinfo{title}{Band-limited angular spectrum
  method for numerical simulation of free-space propagation in far and near
  fields}}.
\newblock {\emph{\JournalTitle{Opt. Express}}} \textbf{\bibinfo{volume}{17}},
  \bibinfo{pages}{19662--19673}, \doiprefix\url{10.1364/OE.17.019662}
  (\bibinfo{year}{2009}).

\bibitem{Matsushima:03}
\bibinfo{author}{Matsushima, K.}, \bibinfo{author}{Schimmel, H.} \&
  \bibinfo{author}{Wyrowski, F.}
\newblock \bibinfo{journal}{\bibinfo{title}{Fast calculation method for optical
  diffraction on tilted planes by use of the angular spectrum of plane waves}}.
\newblock {\emph{\JournalTitle{J. Opt. Soc. Am. A}}}
  \textbf{\bibinfo{volume}{20}}, \bibinfo{pages}{1755--1762},
  \doiprefix\url{10.1364/JOSAA.20.001755} (\bibinfo{year}{2003}).

\end{thebibliography}



\section*{Acknowledgements} 
Giuseppe Calafiore, Heeyoon Lee, and Alexander Koshelev have contributed to the design and fabrication of the waveguide used in this work. 
Clinton Smith designed the compact prototype and provided engineering supports. 
We also thank for valuable discussion provided by Grace Kuo and Suyeon Choi.

\section*{Author contributions statement}
C.J. and K.B. initiated the project, designed the architecture, conceived algorithms, and conducted the experiments. M.C. conducted the experiments. D.L. and B.L. advised and supervised the project. C.J. wrote the initial draft of the manuscript and all authors reviewed the manuscript. 

\section*{Additional information}
\textbf{Accession codes} The source code will be provided upon acceptance;
\textbf{Competing interests} Authors declare no conflicting interest;




\end{document}